\documentclass[10pt]{iopart}
\usepackage{graphicx,epsfig}
\usepackage{bm}
\usepackage{iopams}
\newcommand{\ba}{\begin{eqnarray}}
\newcommand{\ea}{\end{eqnarray}}
\newcommand{\bse}{\numparts}
\newcommand{\ese}{\endnumparts}

\newcommand{\M}{{\cal {M}}}

\newcommand{\bbq}{\begin{quote}}
\newcommand{\eeq}{\end{quote}}
\newcommand{\tbb}{t_{\textrm{\tiny{bb}}}}
\newcommand{\tcoll}{t_{\textrm{\tiny{coll}}}}
\newcommand{\tmax}{t_{\textrm{\tiny{max}}}}
\newcommand{\Lmax}{L_{\textrm{\tiny{max}}}}
\newcommand{\RR}{{}^3{\cal{R}}}

\newcommand{\T}{{}^3{\cal{T}}}

\newcommand{\EE}{{\cal{E}}}
\newcommand{\FF}{{\cal{F}}}

\newcommand{\VV}{{\cal{V}}}
\newcommand{\HH}{{\cal{H}}}

\newcommand{\hOm}{\hat\Omega}

\newcommand{\Dih}{\delta_i^{(\HH)}}
\newcommand{\Dim}{\delta_i^{(m)}}
\newcommand{\Dik}{\delta_i^{(k)}}

\newcommand{\Da}{\delta^{(A)}}

\newcommand{\Dh}{\delta^{(\HH)}}

\newcommand{\Dm}{\delta^{(m)}}

\newcommand{\Dk}{\delta^{(k)}}

\newcommand{\dd}{{\rm{d}}}
\newcommand{\rtv}{r_{\rm{tv}}}

\begin{document}

\title[Radial asymptotics of LTB dust models.]{Radial asymptotics of Lema\^{\i}tre--Tolman--Bondi dust models. } 
\author{ Roberto A. Sussman$^\ddagger$}
\address{
$^\ddagger$Instituto de Ciencias Nucleares, Universidad Nacional Aut\'onoma de M\'exico (ICN-UNAM),
A. P. 70--543, 04510 M\'exico D. F., M\'exico. }
\ead{sussman@nucleares.unam.mx}
\date{\today}
\begin{abstract} We examine the radial asymptotic behavior of spherically symmetric Lema\^{\i}tre--Tolman--Bondi dust models by looking at their covariant scalars along radial rays, which are spacelike geodesics parametrized by proper length $\ell$, orthogonal to the 4--velocity and to the orbits of SO(3). By introducing quasi--local scalars defined as integral  functions along the rays, we obtain a complete and covariant representation of the models, leading to an initial value parametrization in which all scalars can be given by scaling laws depending on two metric scale factors and two basic initial value functions.  Considering regular ``open'' LTB models whose space slices allow for a diverging $\ell$, we provide the conditions on the radial coordinate so that its asymptotic limit corresponds to the  limit as $\ell\to\infty$. The ``asymptotic state'' is then defined as this limit, together with asymptotic series expansion around it, evaluated for all metric functions,  covariant scalars (local and quasi--local) and their fluctuations. By looking at different sets of initial conditions, we examine and classify the asymptotic states of parabolic, hyperbolic and open elliptic models admitting a symmetry center. We show that in the radial direction the models can be asymptotic to any one of the following spacetimes: FLRW dust cosmologies with zero or negative spatial curvature, sections of Minkowski flat space (including Milne's space), sections of the Schwarzschild--Kruskal manifold or self--similar dust solutions.                             
\end{abstract}
\pacs{98.80.-k, 04.20.-q, 95.36.+x, 95.35.+d}

\maketitle
\section{Introduction.}

The spherically symmetric LTB dust models \cite{LTB} are among the best known and most useful exact solutions of Einstein'e equations. Since they allow us to examine non--linear effects analytically, or at least in a tractable way, there is an extensive literature (see \cite{kras,kras2} for comprehensive reviews) using them,  mostly as models of cosmological inhomogeneities \cite{KH1,KH2,KH3,KH4,ltbstuff}, but also as in other theoretical contexts, such as gravitational collapse and censorship of singularities \cite{sscoll,joshi} or quantum gravity \cite{quantum}. There is also a widespread literature \cite{LTB1,LTB2,LTBkolb,num3,num2,LTBfin} considering LTB models as tools to probe how the cosmic acceleration associated to recent observations can be accounted for inhomogeneities, without introducing an exotic source like dark energy. LTB models are also a standard choice \cite{LTBchin,LTBave1,LTBave2,LTBave3} to apply Buchert's scalar averaging formalism \cite{ave_review}, in which the effects of dark energy could be mimicked by ``back--reaction'' terms (see \cite{celerier} for a review of all this literature). In practically all articles the models are parametrized by a standard set of free functions and analytic solutions (implicit and parametric). 

While the literature is vast and exhaustive, there is still room to explore further development on their theoretical properties (see for example \cite{wainwright} in this context). The present article deals with a theoretical problem that has not been, as far as we are aware, previously examined in the literature, namely: the asymptotic behavior of LTB dust models (through their covariant scalars) in the radial spacelike direction, which can be defined in a covariant manner in terms of spacelike geodesics (radial rays) whose tangent vectors are orthogonal to the 4--velocity and to the orbits of SO(3). In the remaining of this section we explain and summarize the contents of the article.

Basic background on LTB models is provided in section 3: the metric, field equations, classification in kinematic classes: parabolic, hyperbolic and elliptic, as well as a covariant time slicing that defines the space slices $\T[t]$ orthogonal to the 4--velocity field and marked by constant values of $t$. In section 3 we introduce an alternative set of quasi--local scalar variables \cite{suss02,suss08,sussQL,suss09,suss10}, leading to a ``fluid flow'' description of the dynamics of the models that is similar (and equivalent) to the ``1+3'' approach of Ellis, Bruni and coworkers \cite{ellisbruni89,1plus3}, which for LTB models (as with all spherically symmetric spacetimes) reduce to scalar equations \cite{LRS}. As we showed in \cite{sussQL,suss09,suss10}, these variables can also be understood in the framework of a perturbation formalism on a FLRW ``background'' defined by the quasi--local scalars, which satisfy FLRW dynamics, while their fluctuations are gauge invariant and covariant non--linear perturbations. The quasi--local variables and their fluctuations are potentially useful for a numeric approach to LTB models (see \cite{suss09} and sections XI--XIII of \cite{suss10}), but they are also very handy for analytic and qualitative work, since they lead to an initial value parametrization of the models in which all covariant scalars can be given by simple scaling laws that depend on the two metric functions (scaled to an initial $\T[t_i]$) and initial value functions.

In section 4 we discuss the definition and properties of the proper radial length, $\ell$, along the radial rays, which is the affine parameter of these geodesics. Since we need to explore how scalars behave as $\ell\to\infty$, we will only consider LTB models (admitting a symmetry center) in which this limit can be realized, which means``open''  models whose 3--dimensional space slices $\T[t]$ orthogonal to the 4--velocity are homeomorphic (topologically equivalent) to $\mathbb{R}^3$, as $\ell$ is always finite at all $\T[t]$ in ``closed'' elliptic models with slices $\T[t]$ homeomorphic to $\mathbb{S}^3$.

In order to probe the asymptotic  behavior of covariant (local and quasi--local) scalars along the radial  rays, we should (ideally) evaluate these scalars as functions of $\ell$, which without numeric work is practically an impossible task (even qualitatively), as $\ell$ is an integral of a metric function that can only be determined (at best) in implicit form in terms of another metric function. Instead, we provide the conditions so that this radial asymptotic regime can be examined in terms of the dependence of scalars on a well defined radial coordinate. Once this is done, the limit $r\to\infty$ will correspond to $\ell\to\infty$, and as a consequence, a covariant characterization of the asymptotic regime can be  provided in terms of the radial coordinate and initial value functions. 

Since we are considering LTB models that comply with basic regularity conditions (no shell crossings and scalars may only diverge at a central singularity), we define in section 5 a regular asymptotic regime in the radial direction by demanding that all scalars are smooth and finite as $r\to\infty$ (which now corresponds to $\ell\to\infty$). We also discuss (Lemmas 1 and 2) the relation between $\ell$ and $R=\sqrt{g_{\theta\theta}}$, which is another important invariant in spherically symmetric spacetimes \cite{hayward}. Since the quasi--local scalars satisfy less complicated scaling laws than local scalars, it is more practical to study their radial asymptotics first. Hence, we prove in Lemma 3 in section 6 that both types of scalars, the local and quasi--local, share the same asymptotic behavior (given a common set of assumptions on the initial value functions). We also choose a radial coordinate gauge (as there is a radial coordinate gauge freedom in the LTB metric).    

In section 7 we consider the radial asymptotic behavior of the $r$--dependent initial value functions, which are the quasi--local scalars and their fluctuations evaluated at a fiducial ``initial'' slice $\T_i=\T[t_i]$ (the $A_{qi}$ and $\Da_i$), and are necessarily restricted by regularity conditions (the Hellaby--Lake conditions \cite{ltbstuff,suss10,ltbstuff1}). We assume for these functions  a uniform asymptotic convergence to specific (but not restrictive) asymptotic trial analytic functions of $r$: power law, logarithmic or exponential. However, before using these convergence forms to probe the scalars $A_q,\,A$ and the $\Da$ by means of the scaling laws derived in section 4, we introduce in section 8 the distinction between the ``asymptotic limit'', which is simply the limit of scalars as $r\to\infty$, and the notion of ``asymptotic state'', which we define as the set of these limits, together with  suitable series expansions around them, evaluated for the metric functions and all covariant scalars and fluctuations.   
 
We examine the asymptotic limits and states separately for parabolic (section 9), hyperbolic (section 10), elliptic (section 11) models and in section 12 for special models with a simultaneous big bang (initial central singularity) and maximal expansion. A summary of these asymptotic limits and states, as well as a brief discussion, are provided in section 13. Depending on the initial value functions, the scalars in open LTB models have as asymptotic limit, either a FLRW cosmology (with zero or negative spatial curvature) or a section of Minkowski spacetime given in ``non--standard'' coordinates that generalize those defining Milne spacetime (the particular solution ``[s2]'' of \cite{ltbstuff}). Within those LTB models whose asymptotic limit is a Minkowski section we can recognize asymptotic states that clearly identify various particular cases of LTB models: sections of Minkowski spacetime that include the Milne universe,  self--similar LTB solutions (see pages 344--345 of \cite{kras2} and \cite{sscoll,selfsim}) or sections of the Schwarzschild--Kruskal spacetime given in terms of coordinates constructed with radial geodesics (Lema\^\i tre and Novikov coordinates, see page 332 of  \cite{kras2}). 

As we comment in section 13, LTB models asymptotic to a FLRW cosmology can be understood in the context of embedding a dust inhomogeneity in a dust FLRW background, without resorting to an artificial matching at a fixed comoving boundary. In other words, these models can be fully relativistic and less artificial representations of the Newtonian ``spherical collapse models'' used as toy models of structure formation (see \cite{suss09} for discussion on this point). This embedding into homogeneous cosmology is not compatible with LTB models asymptotic to Minkowski (in any of their asymptotic states). Instead, some of these models (specially the vacuum dominated hyperbolic ones) can be considered as toy models of a dust inhomogeneity surrounded by a large cosmic void. In this context, these configurations can approximate a spherical realization of the notion of ``finite infinity'' (or ``fi'') suggested by Ellis \cite{fiEllis}, and considered further by Wiltshire \cite{fiWilt}, as a description of an intermediate scale in which galactic cluster structures can be studied as (approximately) asymptotically flat configurations.

The article contains three appendices: Appendix A provides the analytic (parametric and implicit) solutions in the conventional variables, Appendix B discusses the possibility of considering local scalars (instead of quasi--local ones) as initial value functions, while Appendix C summarizes the various particular case spacetimes that follow from specializing the free parameters of LTB models (in our description: the initial value functions). As shown in sections 9--12, the radial asymptotic state of any open LTB model corresponds to a specific spacetime in this list.   
  
\section{LTB models, kinematic classes and a fluid flow time slicing.}

LTB dust models in their conventional variables are characterized by the following metric and field equations $G^{ab}=\kappa\rho u^au^b$  
\begin{equation} ds^2=-c^2dt^2+\frac{R'{}^2}{1+E}\,dr^2+R^2(d\theta^2+ 
\sin^2\theta d\varphi^2),\label{LTB1} \end{equation}
\ba \dot R^2 &=& \frac{2M}{R} +E,\label{fieldeq1}\\
 2M' &=& \kappa \rho \,R^2 \,R',\label{fieldeq2}\ea
where $\rho$ is the rest--mass density, $\kappa=8\pi G/c^2$,\, $E=E(r),\,M=M(r)$, while $\dot R=u^a\nabla_a R=\partial R/\partial (ct)$ and $R'=\partial R/\partial r$. 

It is common usage in the literature  (see \cite{kras,kras2,ltbstuff}) to classify the solutions of (\ref{fieldeq1}) in ``kinematic equivalence classes'' given by the sign of $E$, which determines the existence of a zero of $\dot R^2$. Since $E=E(r)$, the sign of this function can be, either the same in the full range of $r$, in which case we have LTB models of a given kinematic class, or it can change sign in specific ranges of $r$, defining LTB models with regions of different kinematic class (see \cite{ltbstuff}). These kinematic classes are 
\bse\ba E = 0,\qquad \hbox{Parabolic models or regions}\label{par}\\
 E \geq 0,\qquad \hbox{Hyperbolic models or regions}\label{hyp}\\
  E \leq  0,\qquad \hbox{Elliptic models or regions}\label{ell} \ea\ese
where the equal sign in (\ref{hyp}) and (\ref{ell}) holds only in a symmetry center. The solutions of the Friedman--like field equation (\ref{fieldeq1}) for each kinematic class are given in Appendix A.  The case $M=0$ with $E>0$ arbitrary has been classified in \cite{ltbstuff} as the solution ``[s2]'' and is locally equivalent to Minkowski spacetime.  

\subsection{A fluid flow time slicing and covariant scalars.}

The normal geodesic 4--velocity in (\ref{LTB1}) defines a natural time slicing in which the space slices are the 3--dimensional Riemannian hypersurfaces $\T[t]$, orthogonal to $u^a$, with metric $h_{ab}=u_au_b+g_{ab}$, and marked by arbitrary constant values of $t$. 
Each spacelike slice is a warped product $\T[t]=\chi[t](r)\times_R \mathbb{S}^2(\theta,\varphi)$, where the warping function is $R(t,r)\geq 0$, the fibers are concentric 2--spheres $\mathbb{S}^2$ with surface area $4\pi R^2(t,r)$ (orbits of SO(3)), while the leaves $\chi[t](r)$ are ``radial rays'' or curves of the form $C(r)=[ct_0, r,\theta_0,\varphi_0]$, with $t_0,\theta_0,\varphi_0$ constants. The rays are orthogonal to the fibers and isometric to each other in any given $\T[t]$, and are also geodesics of the $\T[t]$ and spacelike geodesics of the LTB metric (\ref{LTB1}). Since we are assuming the existence of (at least) one symmetry center, then every $\chi[t](r)$ for $t$ constant is diffeomorphic to $\mathbb{R}^+=\{x\,|\,x\geq 0\}$. Hence, we will consider every LTB scalar function as equivalent, under the time slicing given by $u^a$, to a one parameter family of real valued functions $A[t]: \mathbb{R}^+\to \mathbb{R}$ so that $A[t](r)=A(t,r)$.

Besides $\rho$ given by (\ref{fieldeq2}), other covariant objects associated with LTB models  are the expansion scalar, $\Theta$, the Ricci scalar of the space slices, $\RR$, plus the shear and electric Weyl tensors, $\sigma_{ab},\,E_{ab}$
\ba \fl\Theta &=& \tilde\nabla_au^a=\frac{2\dot R}{R}+\frac{\dot R'}{R'},\qquad
\RR = -\frac{2(E\,R)'}{R^2R'},\label{ThetaRR}\\
\fl \sigma_{ab} &=& \tilde\nabla_{(a}u_{b)}-(\Theta/3)h_{ab}=\Sigma\,\Xi^{ab},\qquad
E^{ab}=  u_cu_d C^{abcd}=\EE\,\Xi^{ab},\label{SigEE}\ea
where $h_{ab}=u_au_b-g_{ab}$,\, $\tilde\nabla_a = h_a^b\nabla_b$,\, and $C^{abcd}$ is the Weyl tensor, $\Xi^{ab}=h^{ab}-3\eta^a\eta^b$ with $\eta^a=\sqrt{h^{rr}}\delta^a_r$ being the unit tangent vector along the radial rays (orthogonal to $u^a$ and to the orbits of SO(3)).  The scalars $\EE$ and $\Sigma$ in (\ref{SigEE}) are
\begin{equation}\Sigma = \frac{1}{3}\left[\frac{\dot R}{R}-\frac{\dot R'}{R'}\right],\qquad
\EE = -\frac{\kappa}{6}\,\rho+ \frac{M}{R^3}.\label{SigEE1}\end{equation}
Since LTB models (as all spherically symmetric spacetimes) are LRS (locally rotationally symmetric), they can be completely characterized by covariant scalars. Considering (\ref{fieldeq2}), (\ref{ThetaRR}), (\ref{SigEE}) and (\ref{SigEE1}), these are the local ``fluid flow'' scalars  
\begin{equation} \{\rho,\,\Theta,\, \RR,\,\Sigma,\, \EE\},\label{locscals}\end{equation}
whose evolution equations completely determine the dynamics of LTB models in the fluid flow or ``1+3'' approach \cite{ellisbruni89,1plus3}, and thus provide an alternative approach to that based on the analytic solutions of (\ref{fieldeq1}). 

\section{Quasi--local scalars and their fluctuations.}

For every scalar function $A$ in LTB models we define its quasi--local dual $A_q$ as the family of real valued functions $A_q[t]: \mathbb{R}^+\to \mathbb{R}$ given by
\footnote{This section provides the minimal background material to make this article as self--contained as possible. The reader is advised to consult reference \cite{suss10} for details on the quasi--local scalar representation of LTB models.}
\footnote{Since it is clear that $t$ is a fixed arbitrary parameter we will omit henceforth the notation $[t]$ unless it is needed.}
\begin{equation}  A_q=\frac{\int_0^{r}{A\,\FF\,\dd\VV_p}}{\int_0^{r}{\FF\,\dd\VV_p}}=\frac{\int_0^{r}{A R^2 R'\,\dd x}}{\int_0^{r}{R^2 R'\dd x}},\label{aveq_def}\end{equation}
where the integration is along arbirtary slices $\T[t]$,\,with $\FF\equiv (1+E)^{1/2}$,\, $\dd\VV_p=\sqrt{h_{ab}}\dd r\dd\theta\dd\varphi$, and we are using the notation $\int_0^r{... \dd x}=\int_{x=0}^{x=r}{... \dd x}$. The quasi--local scalars comply with the following properties
\bse\ba    
 A_q'=(A_q)' = \frac{3R'}{R}\,\left[A-A_q\,\right],\label{propq2}\\
A(r) - A_q(r) = \frac{1}{R^3(r)}\int_0^r{A' \,R^3 \,\dd x}.\label{propq3}\ea\ese
Given the pair of scalars $\{A,\,A_q\}$, we define the relative fluctuations as
\begin{equation} \Da \equiv \frac{A-A_q}{A_q}=\frac{A'_q/A_q}{3R'/R} =\frac{1}{A_q(r) R^3(r)}\int_0^r{A'\,R^3\,\dd x},\label{Dadef}\end{equation}
where we used (\ref{propq2}) and (\ref{propq3}).

\subsection{Scaling laws and initial value functions.}

The quasi--local scalars lead in a natural manner to an initial value parametrization of LTB models, so that all quantities can be scaled in terms of their value at a fiducial (or ``initial'') slice $\T_i\equiv \T[t_i]$, where $t=t_i$ is arbitrary. Hence, the subindex ${}_i$ will denote henceforth ``initial value functions'', which will be understood to be scalar functions evaluated at $t=t_i$. This procedure suggests rephrasing the metric functions $R$ and $R'$ as dimensionless scale factors 
\begin{equation} L = \frac{R}{R_i},\qquad
 \Gamma = \frac{R'/R}{R'_i/R_i}=1+\frac{L'/L}{R'_i/R_i},\label{LGdef}\end{equation}
leading to scaling laws for all covariant scalars, which now become functions of the scale factors $L,\,\Gamma$ and initial value functions. Considering the definition (\ref{aveq_def}), the quasi--local duals of $\rho,\,\Theta$ and $\RR$ depend only on $L$ and initial value functions
\ba m_q = \frac{m_{qi}}{L^3} = \frac{M}{R^3},\label{mq}\\
 k_q = \frac{k_{qi}}{L^2} =-\frac{E}{R^2},\label{kq}\\
 \HH_q^2 = \frac{\dot L^2}{L^2}=\frac{\dot R^2}{R^2}=2m_q-k_q=\frac{2m_{qi}-k_{qi}L}{L^3},\label{Hq}\ea     
where, to simplify the notation, we have introduced (and will use henceforth) the definitions
\begin{equation}\fl 2m\equiv \frac{\kappa}{3}\,\rho,\quad 2m_q\equiv \frac{\kappa}{3}\,\rho_q,\qquad k\equiv \frac{\RR}{6},\quad k_q\equiv \frac{\RR_q}{6},\qquad \HH\equiv\frac{\Theta}{3},\quad \HH_q\equiv\frac{\Theta_q}{3}.\label{mkHdefs}\end{equation}
Scaling laws for the local scalars follow readily from (\ref{fieldeq2}) and (\ref{ThetaRR}) as:
\bse\ba  m=\frac{m_{qi}}{L^3}\,[1+\Dm] =\frac{m_i}{L^3\,\Gamma},\label{slaw1}\\
    k=\frac{k_{qi}}{L^2}\,[1+\Dk] = \frac{k_i}{L^2\,\Gamma}\,\left[1+\frac{\Gamma-1}{3\,(1+\Dik)}\right].\label{slaw2}\ea\ese
Since local fluid flow scalars in (\ref{locscals}) are expressible in terms of $m_q,\,\HH_q$ and $k_q$ and their fluctuations (\ref{Dadef})
\bse\ba  \fl m=m_q\,\left[1+\Dm\right],\quad \HH=\HH_q\,\left[1+\Dh\right],\quad k=k_q\,\left[1+\Dk\right],\label{qltransf}\\ \fl \Sigma = -\HH_q\,\Dh,\quad \EE=-m_q\,
\Dm.\label{qltransf2}\ea\ese 
we get, with the help from (\ref{Dadef}), (\ref{mq}), (\ref{kq}) and (\ref{Hq}), the scaling laws for the fluctuations
\bse\ba 1+\Dm=\frac{1+\Dim}{\Gamma},\label{slawDm}\\
\frac{2}{3}+\Dk = \frac{2/3+\Dik}{\Gamma},\label{slawDk}\ea\ese  
\ba  2\Dh &=& \frac{2m_q\,\Dm-k_q\,\Dk}{2m_q-k_q}=\frac{2m_{qi}\,\Dm-k_{qi}\,L\,\Dk}{2m_{qi}-k_{qi}\,L}\nonumber\\
 &=& \frac{2m_{qi}\,[\Dim+1-\Gamma]-k_{qi}\,L\,[\Dik+\frac{2}{3}(1-\Gamma)]}{[2m_{qi}-k_{qi}L]\,\Gamma},\label{slawDh}\ea
which allow us to obtain scaling laws for the local expansion scalar, $\HH$, and the scalar function associated with the shear tensor, $\Sigma$.

We note that the scalars $m_q,\,\HH_q,\,k_q$ and their fluctuations are covariant objects, as $M,\,E,\,R,\,\dot R=u^a\nabla_a R$ are invariants in spherically spacetimes \cite{hayward}. Since the conventional variables $M$ and $E$ depend only on $r$, it is convenient to define them as initial value functions 
\begin{equation}M = m_{qi} R_i^3,\qquad E = -k_{qi} R_i^2. \label{ME}\end{equation}
Given (\ref{LGdef}) and (\ref{ME}), the LTB metric (\ref{LTB1}) takes the form
\begin{equation} \dd s^2=-c^2\dd t^2+L^2\left[\frac{\Gamma^2\,{R'_i}^2\,\dd r^2}{1-k_{qi}\,R_i^2}+R_i^2\left(\dd\theta^2+\sin^2\theta\dd \phi^2\right)\right],\label{LTB2}\end{equation}
The Friedman--like equation (\ref{fieldeq1}) now takes the form (\ref{Hq}). Its solutions, are equivalent to those of (\ref{fieldeq1}) in Appendix A, but now expressed in terms of $L,\,m_{qi},\,k_{qi}$. These solutions will be given explicitly in sections 9, 10 and 11. We can compute from these solutions a scaling law for $\Gamma$.   

\subsection{Curvature singularities.}

The scaling laws (\ref{mq})--(\ref{slawDh}) clearly indicate the existence of two possible curvature singularities whose coordinate locus is
\bse\ba L(t,r) &=& 0,\qquad \hbox{central singularity}\label{Lzero}\\
 \Gamma(t,r) &=& 0 \qquad \hbox{shell crossing singularity}.\label{Gzero}\ea\ese
so that for reasonable initial value functions (bounded and continuous), all scalars $A_q=m_q,\,k_q,\,\HH_q$ diverge as $L\to 0$, whereas local scalars $A=m,\,k,\,\HH$ can also diverge if $\Gamma\to 0$ (even if $L>0$). Notice that if $\Gamma>0$, then all scalars $A$ and $A_q$ only diverge at the central singularity $L=0$, which is an intrinsice feature of LTB models. However, if $\Gamma\to 0$ for $L>0$, then all the relative fluctuations $\Da$ diverge (with $A_q\ne 0$), so that local scalars $A$ diverge while their quasi--local duals $A_q$ remain bounded. This is an obviously unphysical effect of shell crossings that must be avoided. We will denote by ``regular LTB models'' all configurations for which shell crossing singularities are absent, thus complying with
\begin{equation} \Gamma > 0 \quad\forall\;\; (ct,r)\quad\hbox{such that}\quad L>0 \label{noshxG}\end{equation}
In order to test this regularity condition we need to compute $\Gamma$, which will be done separately for parabolic, hyperbolic and elliptic models in sections 9, 	10 and 11.  
  
\section{A well behaved proper radial length.} 

In order to examine the radial dependence of scalar functions we need to define the ``radial direction'' in precise and covariant terms. There is no inherent covariant meaning in the radial coordinate. In fact, the metrics (\ref{LTB1}), (\ref{LTB2}) and $h_{ab}$  are invariant under arbitrary re--scalings $r=r(\bar r)$, indicating the existence of a coordinate gauge freedom that can always be used, either to simplify computations or to eliminate any initial value function by using it as radial coordinate. Yet, radial rays are geodesic curves whose affine parameter is radial proper length, and so the radial dependence of scalars at every individual $\T[t]$ acquires a covariant meaning by relating it to this parameter. 

The proper radial length along an arbitrary $\T[t]$ can be defined as the function $\ell[t]:\mathbb{R}^+\to \mathbb{R}$ such that
\ba  \ell[t](r)=\int_{0}^{r}{\sqrt{g_{rr}}\,\dd x}=\int_0^r{\frac{R'}{\FF}\,\dd x}=\int_0^r{L\,\Gamma\, \frac{R'_i}{\FF}\,\dd x},\label{elldef}\\
 \hbox{where}\qquad \FF \equiv \sqrt{1+E} =[1-k_{qi} R_i^2]^{1/2},\label{Fdef}\ea
so that $\ell[t](0)=0$ for all $t$. A well behaved proper length must necessarily satisfy $\ell(r)>0$ and $\ell\,'(r)>0$ for $r>0$ at all $\T[t]$, so that $r_2>r_1\,\Leftrightarrow\,\ell(r_2)>\ell(r_1)$ at all $\T[t]$. Considering (\ref{Lzero}) and (\ref{noshxG}), these requirements are satisfied for regular radial rays in regular LTB models if the following regularity condition among initial value function holds  
\begin{equation} \hbox{sign}\,(R'_i(r)) = \hbox{sign}\,\FF(r).\label{RirF}\end{equation}
Thus, if $R_i$ has a zero, it must be a common zero of $\FF$ (and of the same order). While proper length is the natural parameter to characterize  radial dependence in a covariant manner, it is not convenient to use it as a spacetime coordinate because in general:\, $\ell=\ell(t,r)$ (the same remark applies to another important scalar like $R$). Hence, for practical reasons we need to describe the radial dependence of scalars in terms of their  radial coordinate. Since we can always use radial coordinate gauge freedom to fix $R_i$, (\ref{RirF}) can be understood as a consistency condition on the radial coordinate, so that it effectively mirrors the dependence on proper radial length along radial rays.   

Since (\ref{elldef}) is valid for all $\T[t]$, then for regular LTB models ($\Gamma>0$)  the sign of $R'$ is the same as the sign of $R'_i$ for all $r$. Therefore, as long as (\ref{RirF}) holds, the radial coordinate can be used to probe the radial profiles of scalars at an arbitrary $\T[t]$, as these profiles will be qualitatively analogous to those with respect to $\ell$:
\bse\ba   R' = \frac{\partial R}{\partial \ell}\,\ell\,'=\frac{\partial R}{\partial \ell}\,\frac{R'}{\FF}\quad\Rightarrow\quad \hbox{sign}\,(R') = \hbox{sign}\,(\partial R/\partial\ell),\label{Rrxi}\\
A' = \frac{\partial A}{\partial \ell}\,\ell\,'=\frac{\partial A}{\partial \ell}\,\frac{R'}{\FF}\quad\Rightarrow\quad \hbox{sign}\,(A') = \hbox{sign}\,(\partial A/\partial\ell).\label{Arxi} \ea\ese   
While a zero of $A'$ necessarily corresponds to a zero of $\partial A/\partial\ell$ at any individual $\T[t]$, in time dependent scalars $A=A(t,r)$, a zero of $A'$ can arise (in general) at different values of $r$ (or $\ell$) for different $\T[t]$, or it can arise in some of the $\T[t]$ and not in others, all of this without violating (\ref{noshxG}),  (\ref{RirF}) or (\ref{Rrxi})--(\ref{Arxi}). As we mentioned above, the case of $R$ is different: a zero of $R'$ must be common to a zero of $\FF$, and so it necessarily occurs at a fixed $r$ and is common to all $\T$. 

Notice that (\ref{RirF}) could be violated even if shell crossings are absent ((\ref{noshxG}) holds). In these situations there would be a surface layer at the comoving worldline marking the zero of $R'_i$ \cite{ltbstuff,suss02}. It is evident that such a surface layer also implies an ill--defined proper radial length.

\section{A regular asymptotic regime in the radial direction.} 

A radial asymptotic regime is associated with convergence and behavior of metric functions and covariant scalars the limit $\ell\to\infty$ along radial rays in the space slices $\T(t)$. However, given the existence of (at least) one symmetry center, these slices can be homeomorphic to either $\mathbb{R}^3$ (``open'' models) or to  $\mathbb{S}^3$ (``closed'' models), which must be elliptic because of (\ref{RirF}). Since $\ell$ is everywhere finite in closed models, then the limit $\ell\to\infty$ can only be realized in open models, which can be of all kinematic classes (hyperbolic, parabolic or elliptic with $-1<E<0$). We will only consider open models for the remaining of this article.   

Assuming regular open models for which (\ref{noshxG}) and (\ref{RirF}) hold, we will consider the radial asymptotic regime as regular if the scalars $\{m,\,\HH,\,k\}$ and $\{m_q,\,\HH_q,\,k_q\}$ converge to finite values as $\ell\to\infty$.  If we adopt as a criterion to define a curvature singularity that curvature scalars diverge if a given coordinate locus is reached by geodesics in finite affine parameter values, and since $\ell$ is such a parameter, then there would be no singularity (technically speaking) if these scalars diverge as $\ell\to\infty$. Also, it is possible to conceive a situation in which $k,\,\HH$ and their quasi--local duals diverge (even at finite $\ell$) while the densities $m,\,m_q$ remain bounded, so that a singularity does not arise because curvature scalars remain bounded (see \cite{ltbstuff} and Appendix A of \cite{suss10}). While these situations cannot be ruled out,  we will exclude them from consideration and will, nevertheless, assume henceforth that all covariant scalars ($m,\,k,\,\HH$ and their quasi--local duals) may only diverge at a central singularity (\ref{Lzero}), remaining bounded everywhere and also in the limit $\ell\to \infty$. \footnote{The relative fluctuations $\Da$ might diverge for finite $r$ under regular conditions, see Appendix A4 of \cite{suss10}} 

\subsection{Relation between $\ell$ and $R$.} 

Since $R'>0$ holds everywhere in open models and $R$ is an important invariant scalar involved in the definition of $\ell$, it is necessary to examine the relation between the limits $\ell\to\infty$ and $R\to\infty$. We prove now the following

\begin{quote}

\noindent \underline{Lemma 1}.    Let $\ell(r)$ given by (\ref{elldef}) be the proper radial length along an arbitrary $\T[t]$ homeomorphic to $\mathbb{R}^3$ in a regular LTB model, the limit
\begin{equation}
\mathop {\lim }\limits_{\ell  \to \infty } R(\ell ) = \infty ,\label{Lemma1}\end{equation}
holds for all parabolic and hyperbolic models, and for elliptic models in which $\FF=\sqrt{1+E}$ converges to a nonzero constant as $\ell\to\infty$.\\

\noindent \underline{Proof}. As a consequence of (\ref{Rrxi}), we have
\begin{equation}\fl \qquad \FF(r(\ell)) = \frac{\partial R(\ell)}{\partial \ell}\quad\Rightarrow\quad R(\ell)=\int_0^\ell{\FF(\bar\ell) \dd\bar\ell},\label{RintF}\end{equation}
which is valid at each $\T[t]$ separately. Since $\FF=1$ for parabolic models and $\FF\geq 1$ and $\FF'\geq 0$ hold for all regular hyperbolic models, the result follows directly from (\ref{RintF}). For open elliptic models, $\FF$ lies in the range $0<\FF\leq 1$ with $\FF(0)=1$. If $\FF'\leq 0$ then as $\ell\to\infty$ we have $\FF\to \FF_0$ with $0\leq \FF_0<1$, then (\ref{RintF}) implies
\begin{equation} \FF_0\,\ell\leq R(\ell)\leq \ell,\label{F0}\end{equation}
and so the result follows. If $\FF\to 1$ as $\ell\to\infty$, then $\FF'$ must have a zero at some $r=\rtv$ (which is a minimum of $\FF$ because $\FF(0)=1$), then $\FF\geq \FF(\rtv)$ holds for all $r\ne\rtv$ and $\FF'>0$ for all $r>\rtv$. In this case constraining bounds similar to (\ref{F0}) can be constructed from (\ref{RintF}) with $\FF_0=\FF(\rtv)$ and the result follows. However, if $\FF\to 0$ as $\ell\to\infty$, then (depending on how $\FF$ converges to zero) $R$ might converge to a finite constant in this limit.\\

\noindent \underline{Corollary}. If the limit (\ref{Lemma1}) holds in one $\T[t]$, it must hold in all $\T[t]$. The proof is straightforward, since Lemma 1 is valid for arbitrary $\T[t]$. Also, the $\T[t]$ constitute a smooth foliation of LTB models and (if standard regularity holds) the rays in all $\T[t]$ are complete geodesics in the direction of increasing $\ell$.\\

\noindent \underline{Lemma 2:\, (converse of Lemma 1)}.  The limit
\begin{equation}
\mathop {\lim }\limits_{R  \to \infty } \ell(R ) = \infty \label{Lemma2}\end{equation}
holds for all parabolic and elliptic models, and for all hyperbolic models in which $k_q\to$ constant as $R\to\infty$.\\

\noindent \underline{Proof}. Since $\FF$ is related to $k_q$ by (\ref{kq}) and (\ref{ME}), we can rewrite (\ref{elldef}) (and also (\ref{RintF})) along an arbitrary $\T[t]$ as
\begin{equation} \ell(R)=\int_0^R{\frac{\dd R}{[1-k_q R^2]^{1/2}}},\label{xi_vs_k}\end{equation}    
where we now consider $k_q=k_q(R)$ and we used the fact that $\dd R=R' \dd x$ is an exact differential because $t$ is kept constant as the integral in (\ref{elldef}) is evaluated.   In parabolic models we have $k_q=0$ everywhere, so the result follows trivially from (\ref{xi_vs_k}).  For open elliptic models, (\ref{xi_vs_k}) implies that $k_q R^2\leq 1$ must hold for all $R$, which is equivalent to $1/\FF=[1-k_qR^2]^{-1/2}\geq 1$. Therefore, we have in general
\begin{equation} \ell(R) \geq \int_0^R{\dd R}=R,\label{xi_vs_ke}\end{equation}
and so $\ell\to\infty$ if $R\to\infty$. For hyperbolic models, (\ref{xi_vs_k}) takes the form
\begin{equation} \ell(R)=\int_0^R{\frac{\dd R}{[1+|k_q| R^2]^{1/2}}}.\label{xi_vs_kh}\end{equation}    
It is evident then, that the integral in (\ref{xi_vs_kh}) converges only if $|k_q(R)|$ diverges, hence $\ell(R)$ will always diverge as $R\to\infty$ if $|k_q(R)|\to$ constant in this limit. Notice that this result is valid at individual (but arbitrary) $\T[t]$. In general, $|k_q(R)|$ would converge to a different constant in different $\T[t]$.

\end{quote}

\section{Radial asymptotics of local and quasi--local scalars.} 

Bearing in mind that we are only considering open models for which $R'>0$ holds everywhere, then a convenient (and practical) way to choose the radial coordinate is
\begin{equation}R_i(r)= R_0\,r,\label{Rgauge}\end{equation}
where $R_0$ is a constant length scale. As a consequence of (\ref{Rgauge}), radial dependence becomes dependence on the initial value function $R_i$ and $R_0$ provides a characteristic length scale for the radial coordinate. Since (\ref{Rgauge}) complies with (\ref{RirF}), then $r\to\infty$ implies $\ell\to\infty$ in all open models (including elliptic models for which $R$ tends to a finite constnt in this limit). Also, as a consequence of the corollary of Lemma 1, $r\to\infty$ also implies $R\to \infty$ for all open models (save the above mentioned elliptic models).  Unless specified otherwise, we will henceforth assume that the radial coordinate has been fixed by the gauge (\ref{Rgauge}). 

Assuming the coordinate gauge (\ref{Rgauge}) and that (\ref{noshxG}) and (\ref{RirF}) hold, we examine now the relation between the radial asymptotic behavior of local and quasi--local scalars. 

\begin{quote}

\noindent \underline{Lemma 3}. The following result holds in any given $\T[t]$
\begin{equation}
\mathop {\lim }\limits_{r \to \infty } A(r) = A_0 \quad  \Leftrightarrow \quad \mathop {\lim }\limits_{r \to \infty } A_q (r) = A_0,\label{Lemma3} 
\end{equation}
where $A_0$ is a constant. 

\smallskip

\noindent \underline{Proof}. We consider the case when $A'\geq 0$ for sufficiently large values of $r$. The case when $A'\leq 0$ is analogous. If the limit of $A$ as $r\to\infty$ is $A_0$, then for all $\epsilon>0$ there exists $u(\epsilon)$ such that $A_0-\epsilon<A(r)<A_0$ holds for all $r>u(\epsilon)$. Constraining $A$ by means of this inequality in the definition of  $A_q$ in (\ref{aveq_def}) leads immediately to $A_0-\epsilon<A_q(r)<A_0$, hence $A_q\to A_0$ as $r\to\infty$.  The converse result follows from (\ref{propq2}).

\smallskip

\noindent The following results follow trivially from Lemma 3:

\smallskip

\noindent \underline{Corollary 1}:\, $A_q'\to 0$ and $A'\to 0$ both hold as $r\to\infty$.

\noindent \underline{Corollary 2}:\, $A_q\Da=A-A_q\to 0$ as $r\to\infty$.

\end{quote}

\noindent
Since we have assumed that (\ref{RirF}) holds, these results are valid as $\ell\to\infty$ along individual but arbitrary $\T[t]$, though the constant $A_0$ will be different at different $\T[t]$ and (in general) $A_0=A_0(t)$. These results are useful because it is easier to probe the asymptotics on the quasi--local scalars $A_q$ first, as they satisfy less complicated scaling laws (that do not involve $\Gamma$) and the analytic solutions are given in terms of $m_{qi}$ and $k_{qi}$. Once we have worked out the $A_q$, the asymptotic behavior of the $A$ follows from Lemma 3 and its corollaries.   

\subsection{The relative fluctuations $\Da$.}

\noindent
Lemma 3 yields the following general result:

\begin{quote}

\noindent \underline{Corollary 3 of Lemma 3}:\, If $A\to A_0 \ne 0$ (or, equivalently, $A_q\to A_0 \ne 0$) as $r\to\infty$ in an arbitrary $\T[t]$, then $\Da\to 0$ in this limit, irrespective of how fast or slow $A$ and $A_q$ converge to $A_0$. The proof follows directly from Corollary 2 of Lemma 3.

\end{quote}

\noindent
However, if $A\to 0$ as $r\to\infty$, then this corollary no longer applies. The limit value of $\Da$ in this case is not (necessarily) zero, but depends on the specific asymptotic form in which $A$ and $A_q$ decay to zero.

\section{Asymptotics of initial value functions.} 

The results proven so far are valid at individual (but arbitrary) $\T[t]$. However, it is very difficult to actually test them because we need to evaluate the involved scalars as functions of $r$ for arbitrary fixed $t$. With the exception of parabolic models, this is very hard because $L$ is, in general, given implicitly in the solutions of the Friedman--like equation (\ref{Hq}) (equivalent to (\ref{fieldeq1})). However, the coordinate gauge (\ref{Rgauge}) provides a simple relation between $R_i$ and $r$, and this allows us to obtain general analytic results on the asymptotic behavior of the initial value functions $A_i,\,A_{qi}$ and their fluctuations $\Da_i$. As we show in sections 9, 10, 11 and 12, these analytic expressions yield analytic forms for the asymptotic behavior of these variables for $t\ne t_i$. 

\subsection{Asymptotics of the $\Da_i$}

In order to obtain the limit of the $\Da_i$ when $A,A_q\to A_0=0$, we need to make specific assumptions on the convergence of these scalars to zero.   

\begin{quote} 

\noindent\underline{Uniform asymptotic convergence}. Let $A_i$ and $\tilde A_i$ be smooth integrable scalar functions on $\T[t_i]$ both tending to zero as $r\to\infty$. The scalar $A_i$ uniformly converges asymptotically to $\tilde A_i$ if for every $\epsilon>0$ there exists a real positive number $y$, such that $|A_i-\tilde A_i|<\epsilon$ for all $r>y$. We denote this convergence by $A_i\sim \tilde A_i$. Notice that this definition is also applicable to the quasi--local scalars $A_{qi}$ converging asymptotically to a function $\tilde A_{qi}$. 

\end{quote} 

\noindent
We can assume a given asymptotic convergence either for $A_i$ or for $A_{qi}$. We show bellow how prescribing $A_{qi}$ we obtain $A_i$ and $\Da_i$. The alternative approach (prescribing $A_i$ to obtain $A_{qi}$ and $\Da_i$) is discussed in Appendix B. 

It follows directly from the definition of uniform asymptotic convergence that if $A_{qi}\sim \tilde A_{qi}$ then $A'_{qi}\sim \tilde A'_{qi}$ holds for $r>y$. Therefore, considering (\ref{propq2}) and (\ref{Dadef}), if we assume $A_{qi}\sim \tilde A_{qi}$ we obtain
\bse\ba A_i\sim \tilde A_{qi} + \frac{\tilde A'_{qi}}{3R'_i/R_i}=\tilde A_{qi} +\frac{r\,\tilde A'_{qi}}{3},\label{Asim}\\
\Da_i \sim \frac{\tilde A'_{qi}/\tilde A_{qi}}{3R'_i/R_i}=\frac{\tilde r\,A'_{qi}}{3\,\tilde A_{qi}}.\label{Daasq}\ea\ese
where we eliminated $R_i$ from (\ref{Rgauge}). 

\subsection{Initial density}

While a non--negative $m_{qi}$ is an initial value function of LTB models of all kinematic classes (parabolic, hyperbolic or elliptic), the sign and regularity conditions associated with the quasi--local spatial curvature, $k_{qi}$, depends on the kinematic class. Hence, we examine in this subsection the admissible forms of asymptotic convergence for $m_{qi}$, leaving the discussion for the convergence of $k_{qi}$ for sections 10 and 11 that deal with the asymptotics of hyperbolic and open elliptic models. 

For whatever form we might choose for the asymptotic convergence of $m_i$ or $m_{qi}$, we must be very careful that both of these densities remain non--negative, which (if standard regularity holds and from the scaling laws (\ref{mq}) and (\ref{slaw1})) implies that $m_q$ and $m$ are also non--negative at all $\T[t]$. Since  $m_{qi}$ appears as initial value function in the analytic solutions in our parametrization of the models (see \cite{suss10}), we examine specific asymptotic convergence forms for it. 

Considering (\ref{Asim}), the asymptotic convergence forms for $m_{qi}$ in which both $m_i$ and $m_{qi}$ remain non--negative everywhere are:
\bse\ba   
\fl \hbox{Logarithmic:}\quad\quad m_{qi}\sim m_0/\ln\,r^{\alpha},\;\; (\alpha>0)\quad\qquad M\sim m_0R_0^3 r^3/\ln r^{\alpha}, \label{decmq2}\\ 
\fl \hbox{Power law:}\qquad  m_{qi}\sim m_0 r^{-\alpha}\;\; (0\leq\alpha\leq 3),\qquad\;\; M\sim m_0R_0^3 r^{3-\alpha}.\label{decmq3}\ea\ese
For the power law (\ref{decmq3}) with $\alpha=0$, we have $m_{qi}\sim m_0$, and so Lemma 3 yields $m_i\sim m_0$ and $\Dim\sim 0$. For (\ref{decmq3}) with $\alpha>0$ and for (\ref{decmq2}) $m_{qi}\to 0$, and so Lemma 3 also implies $m_i\to 0$, but the asymptotic behavior of $m_i$ is not identical. We have from (\ref{Asim})--(\ref{Daasq}):
\bse\ba
\fl \hbox{Power law:}\qquad   m_i\sim m_0\left(1-\frac{\alpha}{3}\right) r^{-\alpha},\qquad \Dim \sim \tilde\Dim=-\frac{\alpha}{3}.\label{PL_decq}\\ 
\fl \hbox{Logarithmic:}\qquad m_i\sim \frac{m_0(3\ln r-1)}{3\alpha\ln^2 r},\quad\quad \Dim\sim \tilde\Dim= -\frac{1}{3\ln r}\to 0.\label{Ldecq}
\ea\ese 
Notice that a power law with $\alpha>3$ in (\ref{decmq3}) implies that $m_i$ in (\ref{PL_decq}) becomes negative for sufficiently large $r$, though it remains close to zero and (from Lemma 3) it still tends to zero as $r\to\infty$. The same would happen for an exponential decay $m_{qi}\sim m_0\exp(-\alpha r)$. This undesirable behavior is consistent with the asymptotic forms for $M$ in (\ref{decmq2})--(\ref{decmq3}), since standard regularity requires this function to be monotonously increasing, thus admitting for $m_{qi}$ only a power law decay with $\alpha\leq 3$ or a slow logarithmic decay. The fact that $m_i<0$ happens for steeper decays of $m_{qi}$ follows because $m_i\leq m_{qi}$ holds for $r>y$ if the radial gradients are negative in this range, thus, if $m_{qi}$ becomes very close to zero in a very steep decay it may force $m_i$ to become negative (but close to zero).

While not all assumptions on the decay of $m_{qi}$ to zero yield a positive local density $m_i$, all assumptions on the decay of $m_i\to 0$ necessarily yield $m_i$ and $m_{qi}$ positive in the full range $r>y$. This is so because $m_{qi}\geq m_i\geq 0$ holds in this range (as $m'_i$ and $m'_{qi}$ are negative, see (\ref{propq2})). We discuss this issue in Appendix B. 

\section{Asymptotic limits and asymptotic states.}               

We explore now the asymptotic behavior of regular LTB models along radial rays of arbitrary $\T[t]$ ({\it{i.e.}} $t\ne t_i$ arbitrary and finite), and looking at parabolic, hyperbolic and elliptic models separately. This involves evaluating asymptotic series expansions of the metric functions and covariant scalars around their limit $r\to\infty$, under the assumption that the initial value functions $m_{qi}$ and $k_{qi}$ converge to admissible forms of uniform asymptotic convergence summarized in tables 1 and 2, and given explicitly by (\ref{decmq2})--(\ref{decmq3}) and (see section 10 and 11) (\ref{deckq2})--(\ref{deckq3}) and (\ref{deckq11})--(\ref{deckq22}).  Since every LTB model can be completely determined by $m_{qi}$ and $k_{qi}$ (assuming the gauge (\ref{Rgauge})), once we assume admissible convergence forms for these initial value functions, we can find asymptotic convergence forms and asymptotic series for the metric functions $L,\,\Gamma$, the scalars $m_q,\,\HH_q,\,k_q$, their fluctuations $\Dm,\,\Dh,\,\Dk$, as well as local scalars $m,\,\HH,\,k$ and other auxiliary quantities ($M,\,E,\,\tbb$, etc). All this information completely characterizes an ``asymptotical state'' for every class of models based on a specific assumption on the convergence forms of $m_{qi}$ and $k_{qi}$ and the kinematic class (parabolic, hyperbolic and elliptic). 

The resulting limits and expansions in the characteristic functions contained in the asymptotic states will be compared to the equivalent parameters in those spacetimes, listed in Appendix C, that are particular and limiting cases of LTB models, namely: dust FLRW cosmologies, Schwarzschild--Kruskal, Minkowski (including Milne) and self--similar LTB solutions. The radial asymptotics of specific classes of LTB models can be then characterized on the basis of this comparison, which we express below more precise terms:

\begin{quote}

Let $\{\M,g\}$ be an LTB model manifold with metric $g$ given by (\ref{LTB2}), with $\{\M_{(0)},\bar g_{(0)}\}$ being the spacetime manifold and metric of any one of the particular cases listed in Appendix C. Let $X=\{L,\,\Gamma,\,A,\,A_q\}$, where $A,\,A_q$ denote the scalars $m,\,k,\,\HH$ and $m_q,\,k_q,\,\HH_q$ in $\{\M,g\}$, with $X_{(0)}$ being the equivalent set of functions in $\{\M_{(0)},g_{(0)}\}$. \\

\noindent{\underline{Definition.}} We shall say that $\{\M,g\}$ is {\it radially asymptotic} to $\{\M_{(0)},\bar g_{(0)}\}$, or that $\{\M,g\}$  {\it converges asymptotically} to $\{\M_{(0)},\bar g_{(0)}\}$ in the radial direction, if the series expansion for every function in $X$ around its limit as $r\to\infty$ coincides with its equivalent function in $X_{(0)}$ up to leading terms.  

\end{quote}

\noindent
Since it is clear that we are considering asymptotic behavior along radial rays, we will simply state that a given LTB model or class of models is ``asymptotic to'' or ``converges to'' a given particular case. Notice that it is sufficient to evaluate the limit as $r\to\infty$ for the quasi--local scalars $A_q$, as (from Lemma 3) this limit will be the same for the local scalars $A$. 

However, the evaluation of the strict limit as $r\to\infty$ of the functions in $X$ is not sufficient to characterize the asymptotic behavior of a given class of LTB models, since the same limit can correspond to different asymptotic expansions and convergences. For example, the limit as $r\to\infty$ of the metric (\ref{LTB2}) of some LTB models can coincide with a Minkowski metric in spherical coordinates ($L,\Gamma \to 1$), but if we consider the expansions of $L$ and $\Gamma$ around this limit, the metric of these models could become equivalent (up to leading terms in the expansions) to the metric of an LTB self--similar solution or a section of Schwarzschild--Kruskal spacetime  (whose metrics themselves have a Minkowski limit as $r\to\infty$). In order to distinguish these cases, we need to consider all metric functions and scalars in $X$, evaluate their asymptotic expansions (asymptotic state) and compare them (up to leading terms) with the functions and $X_{(0)}$. Notice that Another point that needs explaining is why we did not include the fluctuations in the set $X$ in the convergence criterion above. The reason is that $A\to 0$ (or equivalently $A_q\to 0$) as $r\to 0$ implies (in general) a nonzero limit for $\Da$, while the equivalent of $\Da$ in $X_{(0)}$ could be identically zero. This can happen when $X_{(0)}$ is associated with a Minkowskian particular case, for which $m=m_q=0$ holds identically, but then we have $m_q\to 0$ and $m\to 0$ in an LTB model converging to this particular case, and thus $\Dm$ does not tend to zero. However, the asymptotic convergence is well characterized because the nonzero limit of $\Dm$ is consistent with $m_q\to 0$ as $r\to\infty$. A summary of the asymptotic behavior for different classes of LTB models obtained in sections 9--12 is given in section 13.         

\section{Radial asymptotics of parabolic models.} 

For $k_{qi}=0$ the Friedmann--like equation (\ref{Hq}) yields a closed analytic expression for $L$
\begin{equation} L =\left[1+\frac{3}{2}\sqrt{2m_{qi}}\,c(t-t_i)\right]^{2/3},\label{par2}\end{equation}
which is equivalent to the parametric solution (\ref{par1}), and we are only considering expanding configurations ($L$ increases for $t>t_i$). The bang time, $\tbb$, follows by setting $L=0$ and $t=\tbb$ in (\ref{par2})
\begin{equation}  c\tbb = ct_i-\frac{2}{3\sqrt{2m_{qi}}}=ct_i-\frac{2}{3\HH_{qi}}.\label{tbbpar}  \end{equation}
By differentiating (\ref{par2}) and from (\ref{LGdef}), we obtain 
\begin{equation} \Gamma = 1+\Dim-\frac{\Dim}{L^{3/2}}.\label{Gp}\end{equation}
Since $k_{qi}=k_i=0$ and $R_i$ has been fixed by (\ref{Rgauge}), the only initial value functions are $m_{qi}$ and $\Dim$. Necessary and sufficient conditions to fulfill (\ref{noshxG}) (absence of shell crossings) are given by the Hellaby--Lake conditions, which for parabolic models and in terms of our initial value functions takes the form \cite{suss10}
\begin{equation} -1\leq \Dim\leq 0,\label{noshx_par}\end{equation}      
which from (\ref{Dadef}) implies $m'_{qi}\leq 0$, so that $m_{qi}$ tends to zero or to a nonzero constant and its admissible asymptotic forms (preventing a negative $m_i$) are given by (\ref{decmq2})--(\ref{decmq3}).  

Assuming an asymptotic convergence $m_{qi}\sim \tilde m_{qi}$ given by either one of  (\ref{decmq2}) or (\ref{decmq3}) the asymptotic convergence for $L$ follows from (\ref{par2}) as
\begin{equation}  L\sim\tilde L=\left[1+\frac{3}{2}\,\sqrt{2\tilde m_{qi}}\,c(t-t_i)\right]^{2/3},\label{asPar1a}\end{equation}
Considering (\ref{mq}), (\ref{Hq}), (\ref{ME}) and (\ref{tbbpar}) we get
\bse\ba
\fl \HH_q\sim \tilde\HH_q=\frac{\sqrt{2\tilde m_{qi}}}{\tilde L^{3/2}},\qquad m_q\sim \tilde m_q=\frac{\tilde m_{qi}}{\tilde L^3}.\label{asPar1b}\\
\fl  M\sim\tilde M=\tilde m_{qi}\,R_0^3\,r^3,\qquad c\tbb\sim ct_i-\frac{2}{3\sqrt{2\tilde m_{qi}}}.\label{asPar1c}\ea\ese
These asymptotic forms are valid for any regular parabolic model, while (from Lemma 3) the limits of $m,\,\HH$ as $r\to \infty$ are the same as those of $m_q,\,\HH_q$. Also, Lemmas 1 and 2 imply that $R\to\infty$ for all regular initial value functions. The asymptotic limits and states depend on the choice of $m_{qi}$. Figures 1a and 1b depict the domain in the $(ct,r)$ plane parabolic models with asymptotic limit to FLRW and Minkowski. 


\subsection{Parabolic models asymptotic to spatially flat FLRW.} 

If $m_{qi}\to m_0>0$, so that $\tilde m_{qi}= m_0$ (the power law form (\ref{decmq3}) with $\alpha=0$), then $\Dim\to 0$ as $r\to \infty$ follows from the corollary 3 of Lemma 1. We have the asymptotic convergence forms
\bse\ba \fl L\sim \tilde L(t)=\left[1+\frac{3}{2}\,\sqrt{2m_0}\,c(t-t_i)\right]^{2/3},\qquad \Gamma\sim 1\label{asPar2a}\\
\fl \HH_q\sim \tilde\HH_q=\frac{\sqrt{2m_0}}{\tilde L^{3/2}(t)},\qquad m_q\sim \tilde m_q=\frac{m_0}{\tilde L^3(t)},\qquad \Dm\sim 0,\label{asPar2c}\\
\fl M\sim \tilde M=m_0\,R_0^3\,r^3,\qquad c\tbb\sim c\tilde\tbb=ct_i-\frac{2}{3\sqrt{2m_0}}=\hbox{const.}.\label{asPar2b}\label{asPar2c}\ea\ese
so that
\begin{equation}\fl L\to \tilde L(t),\quad \Gamma\to 1,\quad m_q\to \tilde m_q(t),\quad \HH_q\to \tilde \HH_q(t),\quad \Dm\to 0,\quad \Dh\to 0.\end{equation}  
Since $\Dim$ and $\Dm$ tend to zero and $\Gamma\to 1$, then $m\to m_q$ and $\HH\to\HH_q$ (which is consistent with Lemma 3). It is evident that the asymptotic state from these expansions is consistent with the corresponding parameters of a spatially flat dust FLRW model (see equations (\ref{FLRW})--(\ref{FLRW_L0})). 

\subsection{Minkowski limit and its asymptotic states.}

If $m_{qi}\sim\tilde m_{qi}\to 0$ as $r\to \infty$, then (\ref{asPar1a}) and (\ref{Gp}) yield 
\begin{equation} \fl L\sim \tilde L\approx 1+\sqrt{2\tilde m_{qi}}\,c(t-t_i),\qquad \Gamma\sim\tilde\Gamma\approx 1+\frac{3}{2}\tilde\Dim \sqrt{2\tilde m_{qi}}\,c(t-t_i).\label{asPar3a}\end{equation}
where we have written ``$\approx$'' (instead of ``$\sim$'') because we used the approximation $(1+\epsilon)^{2/3}\approx 1+(2/3)\epsilon$, valid for $\epsilon\ll 1$.  The power law decay of $m_{qi}$ is useful to illustrate the asymptotic convergence forms for the remaining quantities. Considering (\ref{decmq3}) with $0<\alpha\leq 3$, we obtain
\bse\ba
\fl \HH_q\sim \tilde\HH_q=\frac{\sqrt{2m_0}r^{-\alpha/2}}{\tilde L^{3/2}}\to 0,\qquad m_q\sim \tilde m_q=\frac{m_0\,r^{-\alpha}}{\tilde L^3}\to 0,\label{asPar3c}\\
\fl \Dim\sim\tilde\Dim=-\alpha/3,\qquad \Dm\sim\tilde\Dm= -\alpha/3,\qquad 2\tilde\Dh\approx -\frac{\alpha}{3}.\label{asPar3d}\\
\fl M\sim\tilde M=m_0\,R_0^3\,r^{3-\alpha},\qquad c\tbb\sim ct_i-\frac{2r^{\alpha/2}}{3\sqrt{2m_0}}\to -\infty,\label{asPar3b}\ea\ese
where we used (\ref{slawDm}) and (\ref{slawDh}). From Lemma 3 we have $m\to 0$ and $\HH\to 0$, which agrees with (\ref{qltransf}). If we only consider the limit as $r\to\infty$, then $\tilde m_{qi}\to 0$ clearly implies a Minkowski limit, since $L\to 1$ and $\Gamma\to 1$ transforms the LTB metric (\ref{LTB2}) (with $k_{qi}=0$) into a Minkowski metric in spherical coordinates. The fact that $\Dm,\,\Dh$ tend to a nonzero value is consistent with $m_q,\,\HH_q$ tending to zero (as corollary 3 of Lemma 3 does not apply). If $m_{qi}$ decays logarithmically as in (\ref{decmq2}), we obtain the same Minkowski limit, but with fluctuations $\Da$ now also tending to zero. 

\smallskip

\noindent
The asymptotic states of parabolic LTB models with $\tilde m_{qi}\to 0$ are:  

\begin{itemize}   

\item \underline{Asymptotic to spatially flat self similar solution.} If $\alpha=2$, then $\tilde L=\tilde L(\zeta)$, where $\zeta=c(t-t_i)/r$ is the self--similar variable in (\ref{SSvar}). It is evident that the metric and all functions asymptotically converge to their respective forms in the spatially flat self--similar solution given by (\ref{SS}), (\ref{SS_L0}), (\ref{SSvar})--(\ref{MEtbb42}) with $k_0=0$, which is the case $E=0$ in equation (2.29) of \cite{selfsim} (see also pages 344--345 of \cite{kras2} and \cite{sscoll}). Notice that parabolic LTB models with this asymptotic behavior are not self--similar solutions, they only converge to a self--similar solution as $r\to\infty$ along the $\T[t]$: the latter solution only follows if $m_{qi}=m_0\,r^{-2}$ holds exactly in all the domain of $r$.

\item \underline{Asymptotic to Schwarzschild--Kruskal.} If $\alpha=3$, then a comparison with (\ref{Schw}), (\ref{MEtbb3}) and (\ref{Schw_L0}) implies that we have an asymptotic convergence to Schwarzschild--Kruskal solution in spatially flat comoving coordinates (Lema\^\i tre coordinates, see page 332 of \cite{kras2}). This convergence is consistent with $M\sim m_0R_0^3=$ constant, as opposed to $M$ diverging for $\alpha<3$. Notice that $\Dm\to -1$, hence $m=m_q(1+\Dm)$ converges to zero much faster than $m_q\sim m_0R_0^3/R^3$. 

\item \underline{Asymptotic to Minkowski in generalized Milne coordinates.} If $\alpha\ne 2,3$, then comparison of $\tilde L$ in (\ref{asPar3a}) and (\ref{Mink_L}) shows an asymptotic state compatible with that of a locally Minkowski spacetime in coordinates that generalize Milne's. However, the form (\ref{asPar3a}) does not not follow from the particular solution with $M=m_{qi}=0$ in (\ref{s2case}). Rather, we have an asymptotic state compatible with a section of Minkowski, but one based on expansions around an asymptotic limit $m_{qi}=0$ in models (parabolic) for which $m_{qi}$ is not strictly zero but $k_{qi}=0$ holds everywhere. This Minkowsky asymptotic state occurs also for hyperbolic and elliptic models converging to a parabolic model (see sections 10 and 11).    

\end{itemize} 


\section{Radial asymptotics of hyperbolic models.}

Analytic solutions for hyperbolic models follow from solving the Friedman--like equation (\ref{Hq}) for $k_{qi}<0$. These solutions are equivalent to those of (\ref{fieldeq1}) given by (\ref{hypRt}) and (\ref{hypRt}) in parametric and implicit form. We will work with the implicit form given in terms of our variables by
\begin{equation} \phi = Z_h(x_iL),\label{phi_L_h}\end{equation}
where $Z_h$ is the function 
\begin{equation} u  \mapsto Z_h(u)=u ^{1/2} \left( {2 + u } \right)^{1/2}  - \hbox{arccosh}(1 + u ).\label{hypZ1a}  \end{equation}
and
\bse\ba\phi &\equiv& \phi(t,x_i,y_i) = y_i\, c(t-t_i)+Z_h(x_i),\label{phih}\\
 x_i &=& \frac{|k_{qi}|}{m_{qi}},\qquad y_i=\frac{|k_{qi}|^{3/2}}{m_{qi}}.\label{xy}
\ea\ese
The bang time emerges from setting $L=0$ in (\ref{phi_L_h}) as the following function of $m_{qi}$ and $|k_{qi}|$:
\begin{equation} c\tbb = ct_i-\frac{Z_h(x_i)}{y_i}.\label{tbbh} \end{equation}
The metric function $\Gamma$ follows from (\ref{phi_L_h}) by implicit derivation since it is related to $L'/L$ by (\ref{LGdef}). The result is
\begin{equation}\fl  \Gamma = 1+3(\Dim-\Dik)\left(1-\frac{\HH_q}{\HH_{qi}}\right)-3\HH_q\,c(t-t_i)\,\left(\Dim-\frac{3}{2}\Dik\right),\label{Ghe}\end{equation}
where $\HH_q$ and $\HH_{qi}$ follow from (\ref{Hq}), while $c(t-t_i)$ is given by (\ref{phi_L_h}) and (\ref{phih}). The Hellaby--Lake conditions \cite{ltbstuff,suss10,ltbstuff1} to fulfill the condition (\ref{noshxG}) for absence of shell crossings are given in terms of initial value functions by \cite{suss10}
\begin{equation} c\tbb'\leq 0,\qquad \Dik\geq -\frac{2}{3},\qquad \Dim\geq -1,\label{noshxGh}\end{equation}
where 
\begin{equation}\frac{c\tbb'}{3R'_i/R_i} 
= \frac{\Dim-\Dik}{\HH_{qi}}-c(t_i-\tbb)\left(\Dim-\frac{3}{2}\Dik\right),\label{tbbr}\end{equation}
follows by differentiating (\ref{tbbh}). 

\subsection{Initial negative curvature} 

It is evident from (\ref{Dadef}), (\ref{ME})  that the profile and asymptotic convergence of the initial value function $k_{qi}<0$ in a hyperbolic model is restricted by the second regularity condition in (\ref{noshxGh}), which for a negative $k_{qi}$ necessarily implies that $E=\FF^2-1=-k_{qi} R_0^2 r^2$ must be monotonously increasing (and that $k_i<0$). In order to examine the case $k_{qi}<0$, we remark (see (\ref{Dadef})) that 
\begin{equation} \hbox{sign}(\Dk)=\hbox{sign}(k'_q/k_q). \end{equation}
Hence, $\Dk$ has the same sign in the combination $k_q\leq 0,\, k'_q\geq 0$ as in $k_q\geq 0,\, k'_q\leq 0$. As a consequence, if we assume that $k_{qi}< 0$ holds everywhere (with possibly $k_{qi}\to 0$ asymptotically), then we can examine negative spatial curvature by applying (\ref{PL_decq})--(\ref{Ldecq}) directly to $|k_{qi}|$.  Considering all these points together with (\ref{ME}) and (\ref{Rgauge}), the asymptotic forms of $|k_{qi}|$ and $E$ compatible with standard regularity conditions and with a well defined asymptotic radial range ($r\to\infty \Rightarrow \xi_i\to\infty$) are
\bse\ba  
\fl |k_{qi}|\sim k_0/\ln\,r,\qquad\qquad\quad\qquad \tilde\Dik\to 0,\qquad E\sim k_0\,R_0^2\,r^2/\ln\,r, \label{deckq2}\\ 
\fl |k_{qi}|\sim k_0 r^{-\beta}\;\; (0\leq \beta\leq 2),\qquad\quad \tilde\Dik\to -\frac{\beta}{3},\qquad E\sim k_0\,R_0^2\,r^{2-\beta}.\label{deckq3}\ea\ese
where $k_0>0$ is a constant. Notice that $|k_{qi}|\sim k_0$ (or $\beta=2$) allows for both an increasing $k'_{qi}>0$ or decreasing $k'_{qi}<0$ behavior, though (from Lemma 3) $k'_{qi}\to 0$ as $r\to\infty$ and $|k_{qi}|\to k_0$. In both cases (\ref{deckq2}) and (\ref{deckq3}) we have $|k_{qi}|$ decaying to zero. The asymptotic forms of the local curvature $|k_i|$ are analogous to those of $|k_{qi}|$ given above, since the allowed decay forms are not steeper than $r^{-3}$ (see Appendix B).

\begin{table}
\begin{center}
\begin{tabular}{|c| c| c| c| c| c|}  
\hline
\hline
{} &{} &{} &{Section 10.3} &{Section 10.4} &{Section 10.5}
\\
{$\tilde m_{qi}$} &{$|\tilde k_{qi}|$} &{$\tilde x_i$} &{$\tilde x_i\to 0$} &{$\tilde x_i\to\infty$} &{$\tilde x_i\to x_0$} 
\\ 
\hline
\hline
{} &{} &{} &{} &{} &{} 
\\
{PL} &{PL} &{$x_0\, r^{\alpha-\beta}$} &{$0\leq \alpha<\beta\leq 2$} &{$0\leq\beta<\alpha\leq 3$}  &{$0\leq \alpha=\beta\leq 2$} 
\\
{} &{} &{} &{} &{} &{} 
\\
\hline
\hline
{} &{} &{} &{} &{} &{} 
\\
{PL} &{LOG} &{$x_0\, r^\alpha/\ln r^\beta$} &{$\alpha= 0$} &{$0<\alpha\leq 3$}  &{} 
\\
{} &{} &{} &{$\beta>0$} &{$\beta>0$} &{} 
\\
\hline
\hline
{} &{} &{} &{} &{} &{} 
\\
{LOG} &{PL} &{$x_0\, r^{-\beta}\ln r^\alpha$} &{$0<\beta\leq 2$} &{$\beta= 0$}  &{} 
\\
{} &{} &{} &{$\alpha>0$} &{$\alpha>0$} &{} 
\\
\hline
\hline
{} &{} &{} &{} &{} &{} 
\\
{LOG} &{LOG} &{$x_0\,\alpha/\beta$} &{} &{}  &{$\alpha,\beta > 0$} 
\\
{} &{} &{} &{} &{} &{} 
\\
\hline
\hline
\end{tabular}
\end{center}
\caption{{\bf{Combinations of the admissible asymptotic convergence forms for hyperbolic models}}. The table presents the asymptotic convergence of $\tilde x_i=|\tilde k_{qi}|/\tilde m_{qi}$ for the four possible combinations of the forms $\tilde m_{qi}$ and $|\tilde k_{qi}|$ in (\ref{decmq2})--(\ref{decmq3}) and (\ref{deckq2})--(\ref{deckq3}). The terms ``PL'' and ``LOG'' stand for ``power law'' and ``logarithmic'' decay forms, respectively given by (\ref{decmq2}), (\ref{deckq2}) and (\ref{decmq3}), (\ref{deckq3}). The behavior of $\tilde x_i$ examined in sections 10.3--10.5 follows from the restrictions on the parameters $\alpha,\,\beta$, hence the results of these sections can be readily applied to initial value functions having each asymptotic convergence form.}
\label{table1}
\end{table}


\subsection{Asymptotic approximations for the analytic solutions}

There is no closed explicit form for $L$ as (\ref{par2}), hence we must work with the implicit solutions (\ref{phi_L_h}), with the correspondence rule of the function $Z_h$ given by (\ref{hypZ1a}). Since the admissible forms in (\ref{deckq2})--(\ref{deckq3}) imply that $|k_{qi}|$ is bounded in the radial asymptotic regime, then the radial asymptotic behavior of $\phi=Z_h(x_iL)$ depends on the behavior of $x_i$ in this regime, which in turn depends on the choice of initial value functions $m_{qi}$ and $|k_{qi}|$. Since $Z_h$ is a monotonously increasing function of its arguments, then (\ref{phi_L_h}) implies
\bse\ba 
\fl x_i\to 0\quad\quad\Rightarrow\quad\;\; Z_h(x_i)\to 0\quad \Rightarrow\quad \phi\to 0\quad \Rightarrow\quad Z_h(x_iL)\to 0,\label{ashpat1}\\
\fl x_i\to \infty\quad\;\;\Rightarrow\quad Z_h(x_i)\to \infty\quad \Rightarrow\quad\phi\to \infty\quad \Rightarrow\quad Z_h(x_iL)\to \infty,\label{ashpat2}\ea\ese
while, another possibility is given by
\begin{equation}\fl x_i\to x_0=\hbox{const.},\qquad \phi\to |k_{qi}|^{1/2}\,x_0\,c(t-t_i) +Z_h(x_0)\to Z_h(x_0 L).\label{ashpat3}\end{equation}
The three patterns outlined by (\ref{ashpat1})--(\ref{ashpat2}) and (\ref{ashpat3}), and listed in the 4th, 5th and 6th columns of table \ref{table1}, provide all possible combinations of asymptotic behavior of regular hyperbolic models in terms of their initial conditions (through $x_i$). It is possible to apply appropriate approximations for each one of these cases in order to obtain $L$, which follows by inverting (\ref{phi_L_h})
\begin{equation} L = \frac{Z_h^{-1}(\phi)}{x_i}.\label{L_phi_h}\end{equation}
Given an asymptotic form for $L$ obtained in terms of admissible convergence forms of $m_{qi}$ and $|k_{qi}|$ in (\ref{decmq2})--(\ref{decmq3}) and (\ref{deckq2})--(\ref{deckq3}), we can obtain asymptotic convergence (or approximated) forms for the remaining quantities. We examine separately the asymptotic behavior associated with each of the three cases (\ref{ashpat1})--(\ref{ashpat2}) and (\ref{ashpat3}), each one listed in the fourth, fifth and sixth columns of table \ref{table1}. Figures 1a and 1b respectively depict the domain in the $(ct,r)$ plane of hyperbolic models with asymptotic limit to FLRW and Minkowski. 
  
\subsection{Matter dominated asymptotics: convergence to parabolic models. }

If $m_{qi}$ dominates over $|k_{qi}|$ in the limit $r\to \infty$, then $ x_i\to 0$ and we have the situation described by (\ref{ashpat1}). Hence, we can consider the following approximation for $Z_h$ 
\begin{equation} Z_h(u)\approx \frac{\sqrt{2}}{3}\,u^{3/2}-\frac{\sqrt{2}}{20}\,u^{5/2}\quad\hbox{for}\quad u\approx 0,\label{apprhyp1}\end{equation}
where we use the sign ``$\approx$'' to distinguish this approximation from the approximations that follow from assumptions on uniform asymptotic convergence (``$\sim$'') on the initial value functions. Applying (\ref{apprhyp1}) (up to the leading term) to $Z_h(x_i)$ and $Z_h(x_i L)$ in (\ref{phi_L_h}), and considering that $m_{qi}\sim \tilde m_{qi}$ and $|k_{qi}|\sim |\tilde k_{qi}|$, yields
\begin{equation} L \sim \tilde L\approx \left[1+\frac{3}{2}\sqrt{2\tilde m_{qi}}\,c(t-t_i)\right]^{2/3},\label{asHyp11}\end{equation}
which coincides with the form of $\tilde L$ in (\ref{asPar1a}), though now $L$ does not (necessarily) converge uniformly to $\tilde L$, but approximates it up to a certain order related to the convergence of $x_i$ to zero (hence this approximation depends also on assumptions on $|k_{qi}|$, which is strictly zero in parabolic models). Nevertheless, (\ref{asHyp11}) indicates that hyperbolic models with initial conditions in which $x_i\to 0$ converge to a parabolic model. Though, we have now $|k_{qi}|\ne 0$, hence the asymptotic state might not define a general parabolic model and the forms of the involved quantities should not be identical to those of parabolic models in the previous section. In order to discuss these issues, we examine below the two types of asymptotic states associated with (\ref{asHyp11}) that emerge from the form of $\tilde m_{qi}$:

\begin{itemize}

\item \underline{Asymptotic to spatially flat FLRW.}

If $m_{qi}\sim m_0>0$ (the power law form (\ref{decmq3}) with $\alpha=0$, see top entry in the fourth column of table \ref{table1}), then $\Dim\to 0$ and (\ref{asHyp11}) yields $L\approx \tilde L(t)$ with $\tilde L(t)$ given by (\ref{asPar2a}), which is the scale factor of a spatially flat FLRW dust model. The conventional variables $M,\,E$ and $\tbb$ have the asymptotic forms
\begin{equation}\fl M\sim m_0R_0^3r^3,\quad E\sim |\tilde k_{qi}|R_0 r^2,\quad c\tbb \sim ct_i-\frac{2}{3\sqrt{2m_0}}\to \hbox{const}. \label{asHyp12} \end{equation}
where $|\tilde k_{qi}|$ can take the power law form (\ref{deckq3}) with $\beta>0$ or the logarithmic decay (\ref{deckq2}). Since $\tilde L(t)$ is constant along the $\T[t]$ and $|k_{qi}|\to 0$, then the asymptotic forms of the remaining time--dependent scalars follow from series expansions of (\ref{mq}), (\ref{kq}), (\ref{Hq}), (\ref{slawDm}), (\ref{slawDk}), (\ref{slawDh}) and (\ref{Ghe}) around $|k_{qi}|\sim|\tilde k_{qi}|$, leading up to order $\tilde |k_{qi}|$ to
\bse\ba \fl m_q \approx \frac{m_0}{\tilde L^3},\quad |k_q|\approx \frac{|\tilde k_{qi}|}{\tilde L^2}\to 0,\quad \HH_q\approx \frac{\sqrt{2m_0}}{\tilde L^{3/2}}\left(1+\frac{|\tilde k_{qi}|\tilde L}{4m_0}\right)\to \frac{\sqrt{2m_0}}{\tilde L^{3/2}},\label{asHyp13a}\\
\fl \Gamma\approx 1+ \frac{3\,(\tilde L^{5/2}-1)\,|\tilde k_{qi}|\tilde\Dik}{4\,m_0\,\tilde L^{3/2}}\to 1,\quad \Dm\approx -\frac{3(\tilde L^{5/2}-1)\,|\tilde k_{qi}|\tilde\Dik}{4\,m_0\,\tilde L^{3/2}}\to 0,\label{asHyp13b}\\
\fl \Dk\approx \tilde \Dik -\frac{(2+3\tilde\Dik)\,(\tilde L^{5/2}-1)\,|\tilde k_{qi}|\tilde\Dik}{4\,m_0\,\tilde L^{3/2}}\to \tilde \Dik,\label{asHyp13c}\\
\fl 2\Dh \approx \frac{1}{2}\left(1-\frac{3(\tilde L^{5/2}-1)}{2\tilde L^{5/2}}\right)\,\frac{\tilde L\,|\tilde k_{qi}|\tilde\Dik }{m_0}\to 0.\label{asHyp13d}\ea\ese
where $|\tilde k_{qi}|,\,\,\tilde\Dik$ can take any of the admissible asymptotic forms (\ref{deckq2})--(\ref{deckq3}). The fact that, as $r\to\infty$, we have $m_q,\,\HH_q$ tending to time dependent forms, while $\Gamma\to 1$ and $\Dk$ tends to a perturbative term, clearly shows that these hyperbolic models are asymptotic to the spatially flat FLRW dust cosmology. 

\item \underline{Asymptotic limit to Minkowski.}

If $\tilde m_{qi}\to 0$ and $|\tilde k_{qi}|\to 0$, following either power law decays with $0<\alpha<\beta$ or any combination of power law and logarithmic decay, as given by (\ref{decmq2})--(\ref{decmq3}) and (\ref{deckq2})--(\ref{deckq3}), we have the cases listed in the fourth column of table \ref{table1} (excluding the case PL--PL with $\alpha=0$). For all these combinations of $\tilde m_{qi}\to 0$ and $|\tilde k_{qi}|\to 0$ (\ref{asHyp11}) becomes as $r\to\infty$
\begin{equation} \tilde L \approx 1+\sqrt{2\tilde m_{qi}}\,c(t-t_i)\to 1,\label{asHyp14a}\end{equation}
where we used $(1+\epsilon)^{2/3}\approx 1+2/3\epsilon$, which holds for $\epsilon\ll 1$. The conventional variables $M,\,E$ and $\tbb$ take the asymptotic forms 
\begin{equation}\fl M\sim \tilde m_{qi}R_0^3r^3,\quad E\sim |\tilde k_{qi}|R_0 r^2,\quad c\tbb \sim ct_i-\frac{2}{3\sqrt{2\tilde m_{qi}}}\to -\infty.\label{asHyp14b}  \end{equation}
Hyperbolic models with these initial value functions converge to parabolic models whose asymptotic limit is Minkowski (see previous section), but they do not converge to a general parabolic model of this type. This point can be better illustrated by considering a power law decay for both $\tilde m_{qi}$ and $|\tilde k_{qi}|$ as in (\ref{decmq3}) and (\ref{deckq3}) (see the case PL--PL in the top entry of fourth column of table \ref{table1} with $\alpha>0$). Since we must have $\alpha<\beta$ for $x_i\to 0$ to hold, then there cannot be a convergence to parabolic models for which $2<\alpha\leq 3$ holds because regularity conditions imply $\beta\leq 2$.

\smallskip

The scalars $m_q$ and $|k_q|$ and $\HH_q$ take the forms
\bse\ba \fl m_q\approx \frac{\tilde m_{qi}}{\tilde L^3}\to 0,\quad |k_q|\approx \frac{|\tilde k_{qi}|}{\tilde L^2}\to 0,\quad \frac{|k_q|}{m_q}\approx \frac{|\tilde k_{qi}|}{\tilde m_{qi}}\to 0,\label{asHyp15a}\\
\fl \HH_q\approx \frac{\sqrt{2\tilde m_{qi}}}{\tilde L}\,\left(1+\frac{\tilde x_i}{4}\right)\approx \sqrt{2\tilde m_{qi}}\,\left[1-\sqrt{2\tilde m_{qi}}c(t-t_i)+\frac{\tilde x_i}{4}\right] \to 0,\label{asHyp15b}
 \ea\ese 
where $\tilde L$ is given by (\ref{asHyp14a}). Considering the expansions above and only terms linear in $\sqrt{\tilde m_{qi}}$ or $\tilde x_i$, and since $\sqrt{1+\tilde x_i/2}\approx 1+\tilde x_i/4$ holds for $x_i\ll 1$, we obtain up to leading order 
\bse\ba 
\fl \Gamma \approx 1+ \frac{3}{2}\sqrt{2\tilde m_{qi}}\,c(t-t_i)\,\tilde \Dim \to 1,\label{asHyp16a}\\
\fl \Dm \approx \tilde\Dim- \frac{3}{2}\sqrt{2\tilde m_{qi}}\,c(t-t_i)\,\tilde \Dim\,(1+\tilde\Dim)\to \tilde\Dim,\label{asHyp16b}\\
\fl \Dk \approx \tilde\Dik - \frac{1}{2}\sqrt{2\tilde m_{qi}}\,c(t-t_i)\,\tilde\Dim\, (2+3\tilde\Dik) \to \tilde\Dik,\label{asHyp16c}\\
\fl 2\Dh \approx \tilde \Dim - \frac{3}{2}\sqrt{2\tilde m_{qi}}c(t-t_i)\tilde \Dim (1+\tilde\Dim)-(\tilde\Dim-\tilde\Dik)\frac{\tilde x_i}{2}\to \tilde\Dim.\label{asHyp16d}\ea\ese
where the asymptotic convergence form of $\tilde\Dim,\,\tilde\Dik$ depends on the choice of $\tilde m_{qi},\,|\tilde k_{qi}|$ in (\ref{decmq2})--(\ref{decmq3}) and (\ref{deckq2})--(\ref{deckq3}).      

\end{itemize}

\subsection{Vacuum dominated asymptotics.}

Consider now the case when $|k_{qi}|$ dominates over $m_{qi}$ in the limit $r\to \infty$, so that $ x_i\to \infty$ and we have the situation described by (\ref{ashpat2}), with initial value functions listed in the fifth column of table \ref{table1}. We can consider then the following approximation for $Z_h$ 
\begin{equation} Z_h(u)\approx u-\ln(u)\approx u\quad\hbox{for}\quad u\gg 1,\label{apprhyp2}\end{equation}
where, as in the previous subsection, we distinguish this approximation from that following a given assumption of uniform asymptotic convergence on the initial value functions (hence the sign ``$\approx$''). Applying (\ref{apprhyp2}) (up to the leading term) to $Z_h(x_i)$ and $Z_h(x_i L)$ in (\ref{phi_L_h}), and considering that $m_{qi}\sim \tilde m_{qi}$ and $|k_{qi}|\sim |\tilde k_{qi}|$, yields
\begin{equation} L \sim \tilde L\approx 1+|\tilde k_{qi}|^{1/2}\,c(t-t_i),\label{asHyp21}\end{equation}
so that $L$ approximates the solution of the equation $\dot L^2 = |k_{qi}|$, which is the Friedman equation (\ref{Hq}) with $m_q=0$ and $k_q<0$, or equivalently, the solution of (\ref{fieldeq1}) in (\ref{s2case}) with $M=0$ and $E>0$.  From (\ref{Mink})--(\ref{MEtbb2}) in Appendix C2, these LTB models are the solutions [s2] in \cite{ltbstuff}, and are locally Minkowskian (equivalent to sections of Minkowski spacetime parametrized by non--standard coordinates that generalize the Milne universe). Evidently, hyperbolic models characterized by initial value functions complying with $x_i\to \infty$ converge in the radial direction to these Minkowskian sections. Notice that $k_q<0$ does not imply that spacetime curvature is nonzero, as $k_q$ is the quasi--local dual to the spatial curvature ($k$) of the 3--dimensional slices $\T[t]$. The 3--dimensional scalar curvature of these hypersurfaces can be non--trivial, even in a vacuum flat Minkowski space. On the other hand, $m_q$ is associated with a source $m$ appearing in $T^{ab}$ and generating spacetime curvature. If $|k_{qi}|$ dominates over $m_{qi}$ in the asymptotic radial range, then the frame dependent kinematic effects of the spatial curvature play the major role in the behavior of all incumbent scalars in this range. 

The spatial curvature dominated hyperbolic models lead to the following asymptotic states  depending on the convergence of the initial value function $|k_{qi}|$:

\begin{itemize}

\item \underline{Asymptotic to a Milne Universe.} 

If $|k_{qi}|\sim k_0>0$ (which corresponds to the power law form (\ref{deckq3}) with  $\beta=0$), while $\tilde m_{qi}\to 0$ has any of the forms (\ref{decmq2})--(\ref{decmq3}) with $\alpha>0$, then (\ref{asHyp21}) becomes
\begin{equation} L\approx \tilde L(t) = 1+\sqrt{k_0}\,c(t-t_i),\label{asHyp22a}\end{equation}
while the conventional variables take the asymptotic form
\begin{equation}\fl  M\sim \tilde m_{qi}R_0r^3,\qquad E\sim k_0 R_0^2 r^2,\qquad c\tbb \approx ct_i-\frac{1}{\sqrt{k_0}}\to\hbox{const.} \label{asHyp22b} \end{equation}
By comparing with (\ref{Milne})--(\ref{MEtbb2}), it is evident that hyperbolic models with these initial value functions converge to the locally Minkowskian Milne universe.

\smallskip

Bearing in mind that $\tilde L=\tilde L(t)$ and $m_{qi}\to 0$, then the asymptotic forms of the remaining time--dependent scalars follow from series expansions of (\ref{mq}), (\ref{kq}), (\ref{Hq}), (\ref{slawDm}), (\ref{slawDk}), (\ref{slawDh}) and (\ref{Ghe}) around $\tilde m_{qi}=0$. We obtain up to leading order in $\tilde m_{qi}$ the following forms similar to (\ref{asHyp13a})--(\ref{asHyp13d})
\bse\ba \fl m_q \approx \frac{\tilde m_{qi}}{\tilde L^3}\to 0,\quad |k_q|\approx \frac{k_0}{\tilde L^2},\quad \HH_q\approx \frac{\sqrt{k_0}}{\tilde L}\left(1+\frac{\tilde m_{qi}}{k_0\,\tilde L}\right)\to \frac{\sqrt{k_0}}{\tilde L},\label{asHyp23a}\\
\Gamma\approx 1- \frac{3 c(t-t_i)\tilde\Dim\,\tilde m_{qi}^2}{k_0^{3/2}\,[1+\sqrt{k_0} c(t-t_i)]^2}\to 1, \label{asHyp23b}\\
\Dm\approx \tilde\Dim + \frac{3 c(t-t_i)\tilde\Dim\,(1+\tilde\Dim)\,\tilde m_{qi}^2}{k_0^{3/2}\,[1+\sqrt{k_0} c(t-t_i)]^2}\to \tilde\Dim,\label{asHyp23c}\\
 \Dk\approx  \frac{2 c(t-t_i)\tilde\Dim\,\tilde m_{qi}^2}{k_0^{3/2}\,[1+\sqrt{k_0} c(t-t_i)]^2}\to 0,\label{asHyp23d}\\
 2\Dh \approx  \frac{2\tilde\Dim\,\tilde m_{qi}}{k_0\,[1+\sqrt{k_0} c(t-t_i)]}\to 0.\label{asHyp23e}\ea\ese
where $\tilde m_{qi},\,\,\tilde\Dim$ can take any of the admissible asymptotic forms (\ref{decmq2})--(\ref{decmq3}) with $\alpha>0$.  

\item \underline{Asymptotic to a generalized Milne universe.} 

If $x_i$ diverges as $r\to\infty$, but both $\tilde m_{qi},\,|\tilde k_{qi}|$ given by (\ref{decmq2})--(\ref{decmq3}) and (\ref{deckq2})--(\ref{deckq3}) tend to zero in this limit, we have the case listed in the fifth column of table \ref{table1} (of course, excluding the Milne universe discussed above, hence $\alpha,\,\beta>0$ necessarily holds). As a consequence, $L$ takes the form (\ref{asHyp21}), which implies $L\sim\tilde L\to 1$ as $r\to\infty$. The functions $M$ and $E$ take the forms in (\ref{asHyp14b}) with the appropriate values of $\alpha,\,\beta>0$, while $\tbb$ takes the asymptotic form
\begin{equation} c\tbb \approx ct_i - \frac{1}{|\tilde k_{qi}|^{1/2}} \to -\infty,\label{asHyp31}\end{equation}
Hyperbolic models with these initial value functions converge to generalized versions of the Milne universe (case [s2] in \cite{ltbstuff}). The scalars $m_q$ and $k_q$ are given by the same expressions as in (\ref{asHyp15a}), but with $\tilde L$ given by (\ref{asHyp21}) and $\tilde m_{qi},\,|\tilde k_{qi}|$ complying with $\tilde x_i=|\tilde k_{qi}|/\tilde m_{qi}\to\infty$. The scalar $\HH_q$ in (\ref{Hq}) takes the following asymptotic form
\bse\ba
 \fl\tilde \HH_q &\approx& \frac{|\tilde k_{qi}|^{1/2}}{\tilde L}\,\left(1+\frac{2}{\tilde x_i\,\tilde L}\right)^{1/2} \approx \frac{|\tilde k_{qi}|^{1/2}}{\tilde L}\,\left(1+\tilde\chi_i\right)\nonumber\\
\fl  &\approx& |\tilde k_{qi}|^{1/2}\left[1- |\tilde k_{qi}|^{1/2}c(t-t_i)+\chi_i\right] \to 0,\label{asHyp32a}\\
\fl &\hbox{with}&\qquad \chi_i\equiv \frac{1}{\tilde x_i\,\tilde L}\approx \frac{1}{\tilde x_i}. \label{asHyp32b}\ea\ese 
Considering only terms linear in $|\tilde m_{qi}|^{1/2}$ and $\tilde \chi_i$, we obtain up to leading order 
\bse\ba 
\Gamma \approx 1+ \frac{3}{2}c(t-t_i)\,\tilde\Dik\,|\tilde k_{qi}|^{1/2}\to 1, \label{asHyp33a}\\
\Dm \approx \tilde\Dim -\frac{3}{2}c(t-t_i)\,(1+\tilde\Dim)\,\tilde\Dik\,|\tilde k_{qi}|^{1/2}\to \tilde\Dim,\label{asHyp33b}\\
\Dk \approx \tilde\Dik -\frac{3}{2}c(t-t_i)\,(2+3\tilde\Dik)\,\tilde\Dik\,|\tilde k_{qi}|^{1/2}\to \tilde\Dik,\label{asHyp33c}\\
2\Dh \approx \Dk +2(\tilde\Dim-\tilde\Dik)\,\tilde\chi_i.\label{asHyp33d}\ea\ese
where $\tilde\Dim,\,\tilde\Dik$ depend on the choice of $\tilde m_{qi},\,|\tilde k_{qi}|$ in (\ref{decmq2})--(\ref{decmq3}) and (\ref{deckq2})--(\ref{deckq3}). See figure 1b.

\item \underline{Asymptotic to Schwarzschild--Kruskal.} If we choose the power law decay for $\tilde m_{qi}$ in (\ref{decmq3}) with $\alpha=3$, then 
\begin{equation}m_{qi}\sim m_0r^{-3},\qquad M\sim m_0R_0^3=\hbox{const.},\label{asHyp41}\end{equation}
then a comparison with (\ref{Schw}), (\ref{MEtbb3}) and (\ref{Schw_L0}) reveals an asymptotic convergence to Schwarzschild--Kruskal solution in coordinates given by geodesic observers with positive binding energy (see page 332 of \cite{kras2}). Evidently, the hypersurfaces $\T[t]$ of the Schwarzschild--Kruskal spacetime itself, in this specific time slicing, has an asymptotic limit to Minkowski in the radial direction. The asymptotic form for $|\tilde k_{qi}|$ can be any of the admissible forms (\ref{deckq2})--(\ref{deckq3}), but in order to illustrate the asymptotic behavior of the scalars we will also assume a power law decay $|k_{qi}|\sim |\tilde k_{qi}|=k_0\,r^{-\beta}$ with $\beta\leq 2$, so that
\bse\ba\fl\tilde\chi_i\approx \frac{m_0}{k_0}r^{\beta-3}\ll |\tilde k_{qi}|,\qquad \tilde \Dim \sim -1+O(\tilde\chi_i),\quad \tilde\Dik\sim -\frac{\beta}{3}+O(\tilde\chi_i),\label{asHyp42a}\\
\fl E\sim k_0 R_0^2 r^{2-\beta},\qquad c\tbb\approx ct_i -\frac{r^{\beta/2}}{\sqrt{k_0}}\to-\infty. \label{asHyp42b}
\ea\ese
The asymptotic forms for the remaining scalars simply follow from specializing the parameters of (\ref{asHyp32a})--(\ref{asHyp33c}) to this case:
\bse\ba  \HH_q\approx \frac{\sqrt{k_0}}{r^{\beta/2}}\,\left[1-\frac{\sqrt{k_0}\,c(t-t_i)}{r^{\beta/2}}+\frac{m_0}{k_0\,r^{3-\beta}}\right]\to 0,\label{asHyp43a}\\
 \Gamma \approx 1- \frac{\beta\,\sqrt{k_0}\,c(t-t_i)}{2\,r^{\beta/2}}+O(r^{-\beta})\to 1,\label{asHyp43b}\\
\Dm \approx -1 +O(r^{-\beta})\to -1,\label{asHyp43c}\\
\Dk \approx -\frac{\beta}{3}+ \frac{\beta\,(2-\beta)}{6}\frac{c(t-t_i)}{r^{\beta/2}}+O(r^{-\beta}) \to -\frac{\beta}{3},\label{asHyp43d}\\
\Dh \approx \Dk -3\left(1-\frac{\beta}{3}\right)\,\frac{2m_0}{k_0\,r^{3-\beta}}\to -\frac{\beta}{3}.\label{asHyp43e}\ea\ese
Notice that (\ref{asHyp43c}) implies that local density, $m$, decays to zero much faster than $m_q$, as expected in a convergence to Schwarzschild--like vacuum conditions.

\end{itemize}

\subsection{Generic hyperbolic asymptotics.}

We consider now the case (\ref{ashpat3}) when $x_i\to x_0 = $ constant, listed in the sixth column of table \ref{table1}. We will assume the power law decay (\ref{decmq3}) and (\ref{deckq3}) with $\alpha=\beta$, so that density and spatial curvature decay at the same rate (the treatment of the logarithmic decay is analogous). This decay implies $m_{qi}\sim m_0\,r^{-\gamma}$ and $|k_{qi}|\sim k_0\,r^{-\gamma}$, where $0\leq \gamma\leq 2$ (regularity rules out $\gamma>2$). Hence, we obtain from (\ref{phi_L_h}) 
\begin{equation}
\fl \phi \sim y_0\,\xi+Z_h(x_0),\qquad x_i\sim x_0=\frac{k_0}{m_0},\qquad \xi \equiv\frac{c(t-t_i)}{r^{\gamma/2}},\label{asHyp51}\end{equation}
where $y_0=\sqrt{k_0}\,x_0$ and $Z_h$ is given by (\ref{hypZ1a}). Thus, following (\ref{phi_L_h}) and (\ref{hypZ1a}), we have $L\sim \tilde L$ so that (\ref{L_phi_h}) becomes
\bse\ba y_0\,\xi+Z_h(x_0) \sim Z_h(x_0\,\tilde L(\xi)),\label{asHyp52a}\\
\tilde L(\xi)=\frac{1}{x_0}\,Z_h^{-1}(y_0\,\xi+Z_h(x_0)).\label{asHyp52b} \ea\ese 
where we can write $\tilde L=\tilde L(\xi)$ because $Z_h(x_0)$ is a constant. We have then for $\gamma>0$
\begin{equation}\mathop {\lim }\limits_{r \to \infty } \tilde L = \mathop {\lim }\limits_{\xi  \to 0} \tilde L(\xi ) = \frac{1}{{x_0 }}Z_h^{ - 1} (Z_h (x_0 )) = 1.\label{asHyp52c}
\end{equation}
As a consequence of (\ref{asHyp51}), (\ref{asHyp52a})--(\ref{asHyp52b}) and (\ref{asHyp52c}), and bearing in mind that we are considering $t$ constant, the asymptotic form of $\tilde L$ for very large $r$ is follows from a series expansion of $\tilde L(\xi)$ for $\xi\ll 1$ up to order $\xi$ ({\it{i.e.}} order $O(r^{-\gamma/2})$):
\begin{equation} \tilde L(\xi)\approx \tilde L(0)+L_{,\xi}(0)\,\xi = 1+\sqrt{2m_0+k_0}\,\,\xi,\label{asHyp52d}\end{equation}
where we used (\ref{asHyp52c}) and $L_{,\xi}(0)=[\dd L/\dd\xi]_{\xi=0}$  was computed by implicit derivation of (\ref{asHyp52a}). 

The following asymptotic convergence forms follow from (\ref{mq}), (\ref{kq}), (\ref{Hq}), (\ref{ME}) and (\ref{tbbh}):
\bse\ba \fl \HH_q\sim \frac{[2m_0+k_0\,\tilde L]^{1/2}}{r^{\gamma/2}\tilde L^{3/2}},\quad m_q\sim\frac{m_0}{r^{\gamma}\tilde L^3},\quad |k_q|\sim \frac{k_0}{r^{\gamma}\tilde L^2}.\label{asHyp53b}\\
\fl M\sim m_0R_0^3 r^{3-\gamma},\quad E\sim k_0R_0^2 r^{2-\gamma},\quad c\tbb\sim ct_i-\frac{Z_h(x_0)\,r^{\gamma/2}}{y_0},\label{asHyp53a} \ea\ese
Since $\Dim,\,\Dik\to -\gamma/3$ and considering (\ref{asHyp52d}), then (\ref{slawDm}), (\ref{slawDk}), (\ref{Ghe}) and (\ref{asHyp13b}) lead to the following approximations up to order $r^{-\gamma/2}$
\bse\ba  \Gamma \approx 1-\frac{\gamma}{6}\,(2m_0+k_0)^{1/2}\,\xi,\label{asHyp54a}\\
\Dm \approx -\frac{\gamma}{3}\,\left[1-\frac{3-\gamma}{6}\,(2m_0+k_0)^{1/2}\,\xi\right],\label{asHyp54b}\\ 
\Dk \approx -\frac{\gamma}{3}\,\left[1-\frac{2-\gamma}{6}\,(2m_0+k_0)^{1/2}\,\xi\right],\label{asHyp54c}\\
2\Dh\approx -\frac{\gamma}{3}\left[1-\frac{\gamma\,[2m_0(3-\gamma)+k_0(2-\gamma)]}{6\,(2m_0+k_0)^{1/2}}\,\xi\right].\label{asHyp54d}
\ea\ese
The remaining scalars  $m,\,k,\,\HH$ are approximately equal to $m_q,\,k_q,\,\HH_q$ times a factor $1-\gamma/3$. If instead of a power law form we consider a logarithmic as in (\ref{decmq2}) and (\ref{deckq2}), we obtain the same forms, except that $\xi=c(t-t_i)/\ln r$ and $\Dim,\,\Dik\to 0$.   

The asymptotic expressions (\ref{asHyp52d}), (\ref{asHyp53a})--(\ref{asHyp53b}) and (\ref{asHyp54a})--(\ref{asHyp54d}) hold for any regular hyperbolic model for which $x_i=|k_{qi}|/m_{qi}\to x_0=$ constant. Since the free parameter is $\gamma$, we can identify the following asymptotic states:

\begin{itemize}

\item \underline{Asymptotic to negatively curved FLRW.} If $\gamma=0$, then we have $m_{qi}\sim m_0$ and $|k_{qi}|\sim k_0$, as well as $\xi=c(t-t_i)$. Hence, $\tilde L,\,\tilde \HH_q$ and $\tilde m_q$ depend only on $t$, while asymptotically  $M\propto r^3$ and $E\propto r^2$ and $\tbb\to$ constant. Comparing with (\ref{FLRW}), (\ref{MEtbb1}) and (\ref{FLRW_L}), we can see that $L,\,M,\,E$ and $\tbb$ converge to their corresponding forms of a FLRW dust model with negative spatial curvature. 

\item \underline{Asymptotic to generalized Milne.} If $0<\gamma\leq 2$ then $\xi\to 0$ as $r\to\infty$, thus  $m_q,\,|k_q|$ and $\HH_q$ tend to zero, while $\tbb\to-\infty$ and $\Gamma\to 1$. These limits and the expansions around it, together with (\ref{Mink_L}) clearly point out that models with these parameters are asymptotic to a section of Minkowski in generalized Milne coordinates.  

\item \underline{Asymptotic to the self--similar solution with negative spatial curvature.} If $\gamma=2$, then $m_{qi},\,|k_{qi}|\,\propto r^{-2}$ and all quantities in  (\ref{asHyp52d}), (\ref{asHyp53a})--(\ref{asHyp53b}) and (\ref{asHyp54a})--(\ref{asHyp54d}) converge to their respective forms in the self--similar solution with negative spatial curvature in (\ref{SS}) (this is the case $E>0$ in equation (2.29) of \cite{selfsim}). 

\end{itemize}
 
\section{Radial asymptotics of open elliptic models.}

As with the hyperbolic models, we will use the implicit analytic solution of (\ref{Hq}) for $k_{qi}>0$, equivalent to (\ref{elltR}), which in terms of our variables takes the form
\begin{equation} \phi  = \left\{ \begin{array}{l}
 Z_e(x_iL) \qquad\qquad \hbox{expanding phase}\\ 
  \\ 
 2\pi-Z_e(x_iL) \qquad \hbox{collapsing phase}\\ 
 \end{array} \right.\label{ellZ2}\end{equation}
where the expanding and collapsing phases correspond to $\HH_q>0$ and $\HH_q<0$, and $Z_e$ is given by
\begin{equation} u  \mapsto  Z_e(u)= \arccos(1 - u )-u ^{1/2} \left( {2 - u } \right)^{1/2}.\label{ellZ1a}  \end{equation}
with
\bse\ba\phi &\equiv& \phi(t,x_i,y_i) = y_i\, c(t-t_i)+Z_e(x_i),\label{phie}\\
 x_i &=& \frac{k_{qi}}{m_{qi}},\qquad y_i=\frac{k_{qi}^{3/2}}{m_{qi}},\label{xye}
\ea\ese
The scale factor $L$ is restricted by $0<L\leq \Lmax=2/x_i=2m_{qi}/k_{qi}$. The times associated with the initial singularity, the maximal expansion ($\Lmax$) and the collapsing singularity are given by
\begin{equation}\fl  c\tbb = ct_i-\frac{Z_e(x_i)}{y_i},\qquad c\tmax=c\tbb+\frac{\pi}{y_i},\qquad  c\tcoll=c\tbb+\frac{2\pi}{y_i}.\label{tmc}\end{equation}
Notice that (in general) $\tmax=\tmax(r)$ and $\tcoll=\tcoll(r)$, so (like $\tbb(r)$) neither one is (in general) simultaneous (see section 12 for the cases with simultaneous $\tbb$ or $\tmax$). For every comoving observer $r =$ const., the time evolution is contained in the range $\tbb(r)<t<\tcoll(r)$. 

The metric function $\Gamma$ takes the same form as (\ref{Ghe}), with $\HH_q$ given by (\ref{Hq}) with $k_{qi}>0$ and $c(t-t_i)$ follows from (\ref{ellZ2}) and (\ref{phie}). The Hellaby--Lake conditions \cite{ltbstuff,suss10,ltbstuff1} to comply with (\ref{noshxG}) are
\begin{equation} \frac{c\tbb'}{3R'_i/R_i}\leq 0,\qquad  \frac{c\tcoll'}{3R'_i/R_i}\geq 0,\qquad \Dim\geq -1,\label{noshxGe}\end{equation}
where $c\tcoll'$ follows from differentiating $c\tcoll$ in (\ref{tmc}) 
\begin{equation} \frac{c\tcoll'}{3R'_i/R_i} =\left(\Dim-\frac{3}{2}\Dik\right)\,c(\tcoll-\tbb)+\frac{c\tbb'}{3R'_i/R_i},\label{tcollr}\end{equation}
while $c\tbb'$ has the same form as (\ref{tbbr}) with $c\tbb$ given by (\ref{tmc}). 

\subsection{Initial positive curvature.} 

The asymptotic forms for $k_{qi}$ compatible with standard regularity must comply with the constraint $1+E=\FF^2=1-k_{qi} R_0^2 r^2$, and thus $k_{qi}r^2$ cannot diverge as $r\to\infty$.  Hence, the admissible asymptotic forms are 
\bse\ba 
\fl k_{qi}\sim \tilde k_{qi}=k_0 r^{-\beta}\;\; (\beta\geq 2),\qquad \FF^2\sim 1-k_0 R_0^2 r^{2-\beta},\qquad \Dik\to -\frac{\beta}{3},\label{deckq11}\\
\fl k_{qi}\sim \tilde k_{qi}=k_0\,\textrm{e}^{-\beta r}, \qquad\qquad\quad \FF^2\sim 1-k_0 R_0^2 r^2\textrm{e}^{-\beta r},\quad \Dik\to -\frac{\beta}{3}\,r. \label{deckq22}\ea\ese
Notice that the case $k_{qi}\sim k_0$ ({\it{i.e.}} (\ref{deckq11}) with $\beta=0$) is not admissible as an asymptotic form for $r\to\infty$, hence $k_{qi}\to 0$ holds in this limit for all open elliptic models (though $m_{qi}\to m_0 =$ const. is allowed).  We also remark that the logarithmic decay is also ruled out because $k_{qi}$ must decay at least as fast as $r^{-2}$. Also, from (\ref{slaw2}) and (\ref{slawDk}), if $\Dik<-1$ then we have $k<0$ even if $k_q>0$ in all $\T[t]$. This situation occurs if $\beta>3$ in (\ref{deckq11}) and in (\ref{deckq22}) for which $\Dik\to-\infty$ (see Appendix B of \cite{suss10}).

\subsection{Asymptotic approximations for the analytic solutions.}

As with hyperbolic models, there is no closed explicit form for $L$ and we must work with the implicit solutions (\ref{ellZ2})--(\ref{tmc}), with $Z_e$ given by (\ref{ellZ1a}). 
As opposed to the hyperbolic case, we have now two branches, but $\phi=Z_e(x_iL)$ in (\ref{ellZ2}) is a monotonously increasing/decresing function of $x_iL$ in the expanding/colapsing branches, so it has a branched inverse
\begin{equation}x_i L(\phi ) = \left\{ \begin{array}{l}
 Z_e^{ - 1} (\phi ) \\ 
 Z_e^{ - 1} (2\pi  - \phi ) \\ 
 \end{array} \right..\label{L_phi_e}\end{equation}
However, from (\ref{ellZ1a}), (\ref{LTB2}) and (\ref{Rgauge}) for open elliptic models with $k_{qi}>0$ we must have
\begin{equation} \fl 0<x_i\leq 2\qquad\hbox{and}\qquad 1-k_{qi}r^2R_0^2>0\quad\Rightarrow\quad k_{qi}\,r^2<1/R_0^2,\label{ell_restr}\end{equation}
so that $x_i$ is bounded and $k_{qi}r^2$ cannot be an increasing function as $r\to\infty$ (which is consistent with (\ref{deckq11})--(\ref{deckq22})). Hence (as opposed to the hyperbolic case) $\phi$ must remain bounded. 

As a direct consequence of (\ref{deckq11})--(\ref{deckq22}) and (\ref{ell_restr}), the asymptotic convergence $k_{qi}\sim k_0=$ constant is not allowed. Therefore, open elliptic LTB models are incompatible with a positively curved FLRW asymptotic limit. This is not surprising, since these FLRW dust models are necessarily closed (the $\T[t]$ homeomorphic to $\mathbb{S}^3$). 

As with the hyperbolic models, the asymptotic behavior of open elliptic models depends on the initial value functions through $x_i$. If we take into consideration the restrictions (\ref{deckq11})--(\ref{deckq22}) and (\ref{ell_restr}), then (\ref{ellZ2}) and (\ref{L_phi_e}) allow for the following limiting regimes
\begin{equation} x_i\to 0,\qquad x_i\to x_0<2,\qquad x_i\to 2,\label{asepat}\end{equation}
which will be discussed in detail, respectively, in sections 11.3 to 11.5.  The combinations of the admissible asymptotic convergence forms compatible with (\ref{ell_restr}) that correspond to each of these cases are listed in 4th, 5th and 6th columns of table \ref{table2}.  For reasons that will be explained in the following subsections, the collapsing branch will be needed only in the case $x_i\to 2$ in (\ref{asepat}) (see figure 2).

\begin{table}
\begin{center}
\begin{tabular}{|c| c| c| c| c| c|}  
\hline
\hline
{} &{} &{} &{Section 11.3} &{Section 11.4} &{Section 11.5}
\\
{$\tilde m_{qi}$} &{$\tilde k_{qi}$} &{$\tilde x_i$} &{$\tilde x_i\to 0$} &{$\tilde x_i\to x_0<2$} &{$\tilde x_i\to x_0=2$} 
\\ 
\hline
\hline
{} &{} &{} &{} &{} &{} 
\\
{PL} &{PL} &{$x_0\, r^{\alpha-\beta}$} &{$\beta>\alpha$} &{$2\leq\beta=\alpha\leq 3$}  &{$2\leq\beta=\alpha\leq 3$} 
\\
{} &{} &{} &{$0<\alpha\leq 3$} &{} &{$k_0=2m_0$} 
\\
\hline
\hline
{} &{} &{} &{} &{} &{} 
\\
{PL} &{EXP} &{$x_0\, r^\alpha\textrm{e}^{-\beta r}$} &{$\beta>0$} &{}  &{} 
\\
{} &{} &{} &{$0<\alpha\leq 3$} &{} &{} 
\\
\hline
\hline
{} &{} &{} &{} &{} &{} 
\\
{LOG} &{PL} &{$x_0\, r^{-\beta}\ln r^\alpha$} &{$\beta\geq 2$} &{}  &{} 
\\
{} &{} &{} &{$\alpha>0$} &{} &{} 
\\
\hline
\hline
{} &{} &{} &{} &{} &{} 
\\
{LOG} &{EXP} &{$x_0\, \textrm{e}^{-\beta r}\ln r^\alpha$} &{$\beta>0$} &{}  &{} 
\\
{} &{} &{} &{$\alpha>0$} &{} &{} 
\\
\hline
\hline
\end{tabular}
\end{center}
\caption{{\bf{Combinations of the admissible asymptotic convergence forms for open elliptic models}}. The table presents the asymptotic convergence of $\tilde x_i=\tilde k_{qi}/\tilde m_{qi}$ for the four possible combinations of the forms $\tilde m_{qi}$ and $\tilde k_{qi}$ in (\ref{decmq2})--(\ref{decmq3}) and (\ref{deckq11})--(\ref{deckq22}). The terms ``PL'' and ``EXP'' stand for ``power law'' and ``exponential'' decay forms, respectively given by (\ref{decmq3}), (\ref{deckq11}) and (\ref{decmq2}), (\ref{deckq22}). The behavior of $\tilde x_i$ examined in sections 11.3--11.5 follows from the restrictions on the parameters $\alpha,\,\beta$, hence the results of these sections can be readily applied to initial value functions having each asymptotic convergence form.}
\label{table2}
\end{table}


\subsection{Convergence to parabolic models.}

Since $x_i\leq 2$ there is no possibility for $k_{qi}$ to dominate $m_{qi}$ in the limit $r\to \infty$, though the opposite situation is possible and leads to the limit $ x_i\to 0$ corresponding to the first one of the cases in (\ref{asepat}) and listed in the fourth column of table \ref{table2}. As we show further ahead, the collapsing phase of (\ref{ellZ2}) will not be needed (see figure 2), while $Z_e$ is monotonously increasing in the expanding phase. We have then
\begin{equation}
\fl x_i\to 0\quad\quad\Rightarrow\quad\;\; Z_e(x_i)\to 0\quad \Rightarrow\quad \phi\to 0\quad \Rightarrow\quad Z_e(x_iL)\to 0.\label{asepat1}\end{equation}
Hence, we can approximate $Z_e$ in (\ref{ellZ1a}) by 
\begin{equation} Z_e(u)\approx \frac{\sqrt{2}}{3}\,u^{3/2}+\frac{\sqrt{2}}{20}\,u^{5/2}\quad\hbox{for}\quad u\approx 0,\label{apprell1}\end{equation}
which, up to the leading term, coincides with (\ref{apprhyp1}). As in the density dominated hyperbolic case, we assume $m_{qi}\sim \tilde m_{qi}$ and $k_{qi}\sim \tilde k_{qi}$, and apply (\ref{apprell1}) to $Z_e(x_i)$ and $Z_e(x_i L)$ in the expanding branch of (\ref{ellZ2}), leading exactly to the same form for $L\approx \tilde L$ given by (\ref{asHyp11})
\begin{equation} L \sim \tilde L\approx \left[1+\frac{3}{2}\sqrt{2\tilde m_{qi}}\,c(t-t_i)\right]^{2/3},\label{asEll11}\end{equation}
which indicates that open elliptic models with initial conditions in which $x_i\to 0$ are  asymptotic to a parabolic model. 

While open elliptic and hyperbolic models with $x_i\to 0$ have the same radial asymptotic behavior, their time evolution is radically different, as it is restricted by the initial and collapsing singularity (see figure 2). Considering (\ref{apprell1}) and the fact that we are assuming $x_i\to 0$ and $\tilde k_{qi}\to 0$ for all choices in (\ref{deckq11})--(\ref{deckq22}), the maximal expansion and collapse times $c\tmax,\,c\tcoll$ given by (\ref{tmc}) take the asymptotic forms
\bse\ba c\tmax =ct_i+\frac{\pi-Z_e(x_i)}{k_{qi}^{1/2}\,x_i}\approx ct_i-\frac{\sqrt{2}}{3}\,\tilde x_i^{1/2}+\frac{\pi}{\tilde k_{qi}^{1/2}\,\tilde x_i}\to\infty,\label{ctmax1}\\
c\tcoll =ct_i+\frac{2\pi-Z_e(x_i)}{k_{qi}^{1/2}\,\tilde x_i}\approx ct_i-\frac{\sqrt{2}}{3}\,\tilde x_i^{1/2}+\frac{2\pi}{\tilde k_{qi}^{1/2}\,\tilde x_i}\to\infty.\label{ctcoll1}\ea\ese
where $\tilde m_{qi}$ and $\tilde k_{qi}$ take the admissible forms compatible with $\tilde x_i\to 0$. Since the locus of maximal expansion marked by $c\tmax(r)$ corresponds to a maximum $L=\Lmax=2/x_i$, we have $\Lmax\to\infty$ as $r\to\infty$ and so: $R_{\textrm{\tiny{max}}}=R_0r\Lmax\to \infty$ in this limit.   

Since the collapsing phase corresponds to $ct>c\tmax(r)$, the fact that $ct=c\tmax(r)$ diverges as $r\to\infty$ implies that the radial asymptotic range for all hypersurfaces $\T[t]$ occurs within the expanding phase, which justifies the fact that we did not need to consider the collapsing phase of (\ref{ellZ2}) for the study of the radial asymptotics in this case. This is clearly illustrated by figures 2a and 2b. Since open elliptic models have the same asymptotic behavior as in hyperbolic models in which $x_i\to 0$, the same asymptotic limits arise:

\begin{itemize}

\item \underline{Asymptotic to spatially flat FLRW.}

If $m_{qi}\sim m_0>0$ (the power law form (\ref{decmq3}) with $\alpha=0$, see top entry in the fourth column of table 2), then (\ref{asEll11}) yields $L\approx \tilde L(t)$ with $\tilde L(t)$ given by (\ref{asPar2a}), which is the scale factor of a spatially flat FLRW dust model. The conventional variables $M,\,E$ and $\tbb$ have the asymptotic forms as in (\ref{asHyp12}) (with $-\tilde k_{qi}$ instead of $|\tilde k_{qi}|$), while the maximal expansion and collapse times, $c\tmax$ and $c\tcoll$, are given by (\ref{ctmax1}) and (\ref{ctcoll1}) under the specialization (\ref{decmq3}) with $\alpha=0$, so that $x_i\to 0$. The asymptotic forms of the remaining time--dependent scalars are readily computed as in the hyperbolic case. The result is exactly the same forms as in (\ref{asHyp13a})--(\ref{asHyp13d}).  

\item \underline{Asymptotic to Minkowski.}

If $\tilde m_{qi}\to 0$ and $\tilde k_{qi}\to 0$, following either power law decays with $0<\alpha<\beta$ or any combination of power law and exponenetial decay, as given by (\ref{decmq2})--(\ref{decmq3}) and (\ref{deckq11})--(\ref{deckq22}), we have the cases listed in the fourth column of table 2 (excluding the case PL--PL with $\alpha=0$ which corresponds to the spatially flat FLRW case). For all admissible combinations of initial value functions (\ref{asEll11}) becomes as $r\to\infty$ 
\begin{equation} \tilde L \approx 1+\sqrt{2\tilde m_{qi}}\,c(t-t_i)\to 1,\label{asEll14a}\end{equation}
while the conventional variables $M,\,E$ and $c\tbb$ take the same asymptotic forms as (\ref{asHyp14b}), while $c\tmax$ and $c\tcoll$ follow from (\ref{ctmax1})--(\ref{ctcoll1}), and the scalars $m_q$ and $k_q$ have the same forms as in (\ref{asHyp15a}). Considering only terms linear in $\sqrt{\tilde m_{qi}}$ or $\tilde x_i$, we obtain up to leading order
\begin{equation}\fl \HH_q\approx \frac{\sqrt{2\tilde m_{qi}}}{\tilde L}\,\left(1-\frac{\tilde x_i}{4}\right)\approx \sqrt{2\tilde m_{qi}}\,\left[1-\sqrt{2\tilde m_{qi}}c(t-t_i)-\frac{\tilde x_i}{4}\right] \to 0,\label{asHyp15b}\end{equation}
The remaining quantities have the same asymptotic forms as (\ref{asHyp16a})--(\ref{asHyp16d}), but with $\tilde k_{qi}$ and $\tilde\Dik$ now given by (\ref{deckq11})--(\ref{deckq22}).     

\end{itemize}

\subsection{Generic elliptic asymptotics.}

We examine now the second case in (\ref{asepat2}) listed in the fifth column of table \ref{table2}:
\begin{equation}\fl x_i\to x_0=\hbox{const.}<2,\qquad \phi\to k_{qi}^{1/2}\,x_0\,c(t-t_i) +Z_e(x_0)\to Z_e(x_0L),\label{asepat2}\end{equation}
From table 2, the only combination compatible with $x_i\to x_0$ is that in which both, $\tilde m_{qi}$ and $\tilde k_{qi}$, have the power law forms (\ref{decmq3}) and (\ref{deckq11}) with
same exponent: 
\begin{equation} \fl \tilde m_{qi}=m_0r^{-\gamma},\quad \tilde k_{qi}=k_0r^{-\gamma},\quad2\leq \gamma\leq 3,\quad x_0=\frac{k_0}{m_0}<2.\label{initvalsx0}\end{equation}
where the restriction on $\gamma$ follows from the possible common exponents in (\ref{decmq3}) and (\ref{deckq11}). Considering that $Z_e(x_0)<\pi$ for all $x_0<2$ (we examine the case $x_0=2$ in the following subsection), the maximal expansion and collapse times for these forms of $\tilde m_{qi}$ and $\tilde k_{qi}$ take the following asymptotic forms 
\bse \ba c\tmax \sim ct_i+\frac{\pi-Z_e(x_0)}{\sqrt{k_0}\,x_0}\,r^{\gamma/2}\to\infty,\label{ctmax2}\\
c\tcoll \sim ct_i+\frac{2\pi-Z_e(x_0)}{\sqrt{k_0}\,x_0}\,r^{\gamma/2}\to\infty.\label{ctcoll2} \ea\ese
As a consequence,  only the expanding phase in (\ref{ellZ2}) is needed to examine the radial asymptotics in this case (just as with the case $x_i\to 0$). If $x_i\to x_0$, then as $r\to\infty$ the scale factor $L$ at maximal expansion tends to a finite value:\, $L=\Lmax=2/x_i\to 2/x_0$, though we have $R_{\textrm{\tiny{max}}}=R_0r\Lmax\to (2R_0/x_0)\,r\to \infty$ in this limit. Since $2/x_0>1$ for $x_0<2$ and $L=1$ at the initial hypersurface $\T[t_i]$, then this hypersurface necessarily lies in the expanding phase: $t_i<\tmax$ (see figure 2b).  

Since we are only considering the expanding phase in (\ref{ellZ2}), the equation $\phi=Z_e(x_0\tilde L)$ has exactly the same form as in (\ref{asHyp51}) and (\ref{asHyp52a})--(\ref{asHyp52b}), with $Z_e$ instead of $Z_h$. Thus, (\ref{asHyp52c}) holds and we  have $L\to 1$ as $r\to\infty$, while the asymptotic expansion (\ref{asHyp52d}) takes now the form
\begin{equation}\tilde L \approx 1+\sqrt{2m_0-k_0}\,\,\xi,\qquad \xi=\frac{c(t-t_i)}{r^{\gamma/2}}.\label{asEll52d}\end{equation}
The asymptotic convergence forms for $M,\,E,\,\tbb,\,m_q,\,k_q$ and $\HH_q$ follow from (\ref{mq}), (\ref{Hq}), (\ref{ME}) and (\ref{tmc}), and are very similar to those in (\ref{asHyp53a})--(\ref{asHyp53b}):
\bse\ba \fl M\sim m_0R_0^3 r^{3-\gamma},\quad E\sim -k_0R_0^2 r^{2-\gamma},\quad c\tbb\sim ct_i-\frac{Z_e(x_0)\,r^{\gamma/2}}{y_0}\to-\infty,\label{asEll53a}\\
\fl \HH_q\sim \frac{[2m_0-k_0\,\tilde L]^{1/2}}{r^{\gamma/2}\tilde L^{3/2}}\to 0,\quad m_q\sim\frac{m_0}{r^{\gamma}\tilde L^3}\to 0,\quad k_q\sim \frac{k_0}{r^{\gamma}\tilde L^2}\to 0.\label{asEll53b}\ea\ese
where $y_0=\sqrt{k_0}x_0$ and $\tilde L$ is given by (\ref{asEll52d}). 
Since $\Dim,\,\Dik\to -\gamma/3$ and considering (\ref{asEll52d}), then (\ref{slawDm}), (\ref{slawDk}), (\ref{Ghe}) and (\ref{asHyp13b}) lead to the same approximations (up to order $r^{-\gamma/2}$) as (\ref{asHyp54a})--(\ref{asHyp54d}), but with $k_0$ replaced by $-k_0$. Since $\gamma=0$ is not possible, then all open elliptic models complying with (\ref{initvalsx0}) have a Minkowski asymptotic limit. The following asymptotic states emerge:   

\begin{itemize}

\item \underline{Asymptotic to the self--similar solution with positive spatial curvature.} If $\gamma=2$, then $m_{qi},\,k_{qi}\,\propto r^{-2}$ and $M\propto\,r$ and $E\sim -k_0R_0^2$ hold. Hence, there is asymptotic convergence to the self--similar solution with positive spatial curvature in (\ref{SS}) with self--similar variable $\xi=c(t-t_i)/r=\zeta$  (this is the case $E<0$ in equation (2.29) of \cite{selfsim}).

\item \underline{Asymptotic to Schwarzschild--Kruskal.} If $\alpha=3$, then $M\sim m_0R_0^3=$ constant. Comparison with (\ref{Schw}), (\ref{MEtbb3}) and (\ref{Schw_L0}) reveals an asymptotic convergence to Schwarzschild--Kruskal solution in coordinates given by geodesic observers with negative binding energy (Novikov coordinates, see page 332 of \cite{kras2}).

\end{itemize}

\subsection{The case $x_0=2$.}

The particular case $x_i\sim x_0=2$ is characterized by the same initial value functions in (\ref{initvalsx0}) with $k_0=2m_0$ (sixth column of table \ref{table2}). Equation (\ref{ellZ2}) can be rewritten as
\begin{equation} Z_e(2\tilde L(\xi)) \sim \pi \pm 2\sqrt{2m_0}\,\xi,\qquad \xi=\frac{c(t-t_i)}{r^{\gamma/2}}\label{asepat3}\end{equation}
where the $+$ and $-$ signs respectively correspond to the expanding and collapsing phase, and we used the fact that $Z_e(2)=\pi$. 

There is a fundamental difference (in comparison with the case $x_0<2$) in the locus of the maximal expansion time, $c\tmax$. Considering that both $m_{qi}$ and $k_{qi}$ decay as $r^{-\gamma}$, then we can write in general $x_i\sim 2+O(r^{-\gamma})$. Hence, we have for $x_i\approx 2$
\begin{equation}\pi - Z_e(x_i)\approx 2\sqrt{2}r^{-\gamma/2}\,\left[1+\frac{1}{6}\,r^{-\gamma/2}+O(r^{-\gamma})\right],\label{x02exp}\end{equation}
which, from (\ref{tmc}), leads to 
\bse\ba c\tmax \sim ct_i +\frac{[\pi-Z_e(x_i)]\,r^{\gamma/2}}{2\sqrt{2m_0}}\approx ct_i+\frac{1}{\sqrt{m_0}},\label{ctmax02}\\
c\tcoll \sim ct_i +\frac{[2\pi-Z_e(x_i)]\,r^{\gamma/2}}{2\sqrt{2m_0}}\approx ct_i+\frac{\pi\,r^{\gamma/2}}{2\sqrt{2m_0}}\to\infty,\label{ctcoll02}\\
c\tbb \sim ct_i -\frac{Z_e(x_i)\,r^{\gamma/2}}{2\sqrt{2m_0}}\approx ct_i-\frac{\pi\,r^{\gamma/2}}{2\sqrt{2m_0}}\to -\infty,\label{ctbb02}
\ea\ese
so that $\tbb$ and $\tcoll$ tend to curves that are symmetric with respect to the constant asymptotic line of $\tmax$ (see figure 2c). Therefore, as a consequence of (\ref{ctmax02})--(\ref{ctbb02}), there exist an extended collapsing region for open elliptic models for which $x_i\sim x_0=2$. As opposed to the cases $x_i\to 0$ and $x_i\to x_0<2$, there is now a radial asymptotic range for the hypersurfaces $\T[t]$ in both the expanding phase and the collapsing phase $c\tmax<ct<c\tcoll$. In fact, models in which $x_0=2$ are asymptotic to the models with a simultaneous $\tmax$ that will be examined in the following section. 

If $x_i\to 2$, then:\, $L=\Lmax=2/x_i\to 1$ as $r\to\infty$, but  $R_{\textrm{\tiny{max}}}=R_0r\Lmax\to R_0\,r=R_i\to \infty$ in this limit. Notice that  (\ref{ctmax02}) implies that the locus of the maximal expansion itself lies in the expanding phase. The asymptotic limit of $L$ follows by inverting (\ref{asepat3}):
\begin{equation}2\tilde L \sim Z_e^{-1}(\pi\pm 2\sqrt{2m_0}\,\xi).\label{L_phi_x02}\end{equation}
Since $\xi\to 0$ as $r\to\infty$ and $\pi=Z_e(2)$, then (\ref{L_phi_x02}) implies $L\sim \tilde L\to 1$ in this limit (for both the expanding and collapsing phases). While this is the same limit as in the case $x_0<2$, the asymptotic expansions like (\ref{asEll52d}) and the equivalents of (\ref{asHyp54a})--(\ref{asHyp54d}) must be computed up to $\xi^2$, since now $\sqrt{2m_0-k_0}=0$. The result (valid for the expanding and collapsing phases) is
\begin{equation} \tilde L(\xi) \approx \tilde L(0)+\frac{1}{2}\tilde L_{,\xi\xi}(0)\,\xi^2=1-\frac{m_0}{2}\,\xi^2,\label{Laprx02}\end{equation}
where $\tilde L_{,\xi\xi}(0)=[\dd^2\tilde L/\dd\xi^2]_{\xi=0}$ was computed by implicit derivation of (\ref{asepat3}) with respect to $\xi$. The asymptotic form for $\HH_q$ follows readily from (\ref{asEll53b}) with $k_0=2m_0$ and (\ref{Laprx02})
\begin{equation}\fl \HH_q \sim \frac{\sqrt{2m_0}\,(1-\tilde L)^{1/2}}{r^{\gamma/2}\,\tilde L^{3/2}}\approx \frac{m_0\,\xi}{1-3m_0\xi^2/4}\approx m_0\,\xi\,\left[1+\frac{3m_0}{4}\,\xi+O(\xi^2)\right]\to 0.\label{Hqaprx02}\end{equation}
The asymptotic form for $\Gamma$ cannot be computed from (\ref{Ghe}) because $\xi=0$ at $t=t_i$, hence $\HH_{qi}\approx 0$. Instead, we use the definition of $\Gamma$ in  (\ref{LGdef}) applied to $\tilde L$ in (\ref{Laprx02})
\begin{equation} \Gamma \sim 1+\frac{r\tilde L'}{\tilde L} \approx \frac{1+\gamma\,m_0\,\xi^2/2}{1+m_0\,\xi^2/2}\approx 1+\frac{(\gamma-1) m_0}{2}\,\xi^2\to 1. \label{Gaprx02}\end{equation}
Bearing in mind that $\tilde\Dim =\tilde\Dik =-\gamma/3$, the following asymptotic forms follow readily
\bse\ba \Dm \approx -\frac{\gamma}{3}\,\left[1+\frac{(3-\gamma)\,m_0}{2}\,\xi^2+O(\xi^3)\right]\to -\frac{\gamma}{3},\label{Dmaprx02}\\
\Dk \approx -\frac{\gamma}{3}\,\left[1+\frac{(2-\gamma)\,m_0}{2}\,\xi^2+O(\xi^3)\right]\to -\frac{\gamma}{3},\label{Dkaprx02}\\
2\Dh \approx -\frac{2\gamma}{3}\,\left[1+\frac{(2-\gamma)\,m_0}{4}\,\xi^2+O(\xi^3)\right]\to -\frac{2\gamma}{3}.\label{Dhaprx02}\ea\ese
Hence, the case $x_0=2$ yields the same asymptotic states as $x_0<2$, but with a faster decay $O(\xi^2)$ to the Minkowski limit. The models are also asymptotic to Minkowski in generalized Milne coordinates when $2<\gamma<3$, with $\gamma=2,3$ corresponding to models asymptotic to self--similar and Schwarzschild--Kruskal spacetimes. 

\section{Simultaneous big bang or maximal expansion.}

So far we have assumed that the initial curvature singularity ($L=0$) is not simultaneous, but given by the curve $[c\tbb(r),r]$ in the $(ct,r)$ plane. While a constant $\tbb$ is incomplatible with parabolic models, non--trivial and perfectly regular hyperbolic and elliptic models exist for which $\tbb'=0$ (see \cite{KH4,ltbstuff,suss02}). As shown in \cite{suss10}, the Hellaby--Lake conditions to avoid shell crossings are simply (\ref{noshxGh}) and (\ref{noshxGe}) with $\tbb'=0$. Also, the locus $[c\tmax(r),r]$ of the maximal expansion time ($\HH_q = 0$) in elliptic models or regions is, in general, not simultaneous, though regular models exist with $c\tmax'=0$ \cite{KH4,suss10}. We examine in this section the radial asymptotic behavior of regular models with these characteristics.

\subsection{Simultaneous big bang.}   

Let $\tbb=t_{(0)}$ denote the constant time value associated with $L=0$, then from (\ref{tbbh}) and (\ref{tmc}), the initial value functions $m_{qi}$ and $k_{qi}$ are necessarily linked by the constraints
\bse\ba \fl c(t_i-t_{(0)})=F_h(m_{qi},|k_{qi}|)\equiv m_{qi}\frac{Z_h(|k_{qi}|/m_{qi})}{|k_{qi}|^{3/2}},\qquad\hbox{hyperbolic},\label{tbbch}\\
\fl c(t_i-t_{(0)})=F_h(m_{qi},k_{qi})\equiv m_{qi}\frac{Z_e(k_{qi}/m_{qi})}{k_{qi}^{3/2}},\qquad\hbox{elliptic},\label{tbbce}\ea\ese
where the functions $Z_h$ and $Z_e$ are given by (\ref{hypZ1a}) and (\ref{ellZ1a}), and we have used (\ref{phih}) and (\ref{phie}) to express $x_i$ and $y_i$ in terms of $m_{qi}$ and $k_{qi}$. We can prescribe one of the pair $m_{qi}, k_{qi}$ and find the other by solving the constraints (\ref{tbbch}) and (\ref{tbbce}). While these constraints cannot be solved analytically, we examine below the radial asymptotics of models with $\tbb=t_{(0)}$ by looking at them qualitatively. 

Since the functions $m_{qi},k_{qi}$ that solve (\ref{tbbch}) and (\ref{tbbce}) for given values $c(t_i-t_{(0)})$ define the level curves of $F_h(m_{qi},|k_{qi}|)$, we need to verify first if the admissible assumptions of asymptotic convergence of $m_{qi},k_{qi}$ are compatible with these functions lying in a level curve, and then if their limit as $r\to\infty$ (in the $(m_{qi},k_{qi})$ plane) fixes the value of $c(t_i-t_{(0)})$. Considering these assumptions, the following mutually exclusive cases emerge:

\begin{itemize}

\item  $m_{qi}\sim m_0,\,k_{qi}\sim k_0$. These assumptions are only possible in hyperbolic models, hence we have $F_h(m_{qi},|k_{qi}|)\sim F_h(k_0,m_0) =$  constant, which fixes the level curve $c(t_i-t_{(0)})$. Hence, all hyperbolic models with these initial value functions and $\tbb=t_{(0)}$ are asymptotic to negatively curved FLRW dust cosmology, as the latter follows from these asymptotic convergence forms.   

\item $k_{qi}\to 0$ and $m_{qi}\sim m_0$. For models with a simultaneous big bang having these convergence forms we need to find the limit of $F_h$ and $F_e$ as $k_{qi}\to 0$ with $m_{qi}= m_0$ fixed:
\begin{equation}
\mathop {\lim }\limits_{|k_{qi} | \to 0} F_h (|k_{qi} |,m_0 ) = \mathop {\lim }\limits_{k_{qi}  \to 0} F_e (k_{qi} ,m_0 ) = \frac{2}
{{3\sqrt {2m_0 } }}
\end{equation}
which fixes the constant bang time in terms of the level curve $c(t_i-t_{(0)}) = 2/(3\sqrt{2m_0})$. As a consequence, hyperbolic and elliptic models with $\tbb=t_{(0)}$ and these initial value functions are asymptotic to a spatially flat FLRW dust cosmology, as the latter corresponds to these asymptotic convergence forms of $m_{qi}$ and $k_{qi}$.

\item $m_{qi}\to 0$ and $k_{qi}\sim k_0$. This asymptotic convergence only occurs in hyperbolic models ($k_{qi}<0$), since in elliptic models $0<k_{qi}\leq 2m_{qi}$ must hold. To examine this case in models with a simultaneous big bang, we evaluate the limit of $F_h$ as $m_{qi}\to 0$ with $k_{qi}= k_0$ fixed:
\begin{equation}
\mathop {\lim }\limits_{m_{qi}  \to 0} F_e (k_0 ,m_{qi} ) = \frac{1}
{{\sqrt {k_0 } }}
\end{equation}
which fixes the constant bang time in terms of the level curve $c(t_i-t_{(0)}) =1/\sqrt{k_0}$. Thus, hyperbolic models with $\tbb=t_{(0)}$ and these initial value functions are asymptotic to the Milne Universe, which corresponds to these asymptotic convergence forms of $m_{qi}$ and $k_{qi}$.

\item $k_{qi}\to 0$ and $m_{qi}\to 0$. In order to examine this asymptotic convergence for models with a simultaneous big bang, we explore the behavior of $F_h$ and $F_e$ in the direction in the $(m_{qi},k_{qi})$ plane given by the ray $k_{qi}=\alpha_0 m_{qi}$, where $\alpha_0>0$ is a constant ($\alpha_0\leq 2$ for elliptic models). Hence, $F_h$ and $F_e$ along this direction are given by
\begin{equation}F_h(|k_{qi}|)=\frac{\alpha_0\,Z_h(1/\alpha_0)}{|k_{qi}|^{1/2}},\qquad F_e(k_{qi})=\frac{\alpha_0\,Z_e(1/\alpha_0)}{k_{qi}^{1/2}},\end{equation}
Thus, in the limit as $m_{qi}\to 0$ and $k_{qi}\to 0$ (which reduces in this direction to the limit as $k_{qi}$ tends to zero) we have $F_h\to\infty$ and $F_e\to\infty$. In fact,  both $F_h$ and $F_e$ only diverge as $(m_{qi},k_{qi})\to (0,0)$. All this means that $c(t_i-t_{(0)})\to\infty$ for any choice of $m_{qi},k_{qi}$ in which both tend to zero asymptotically. As a consequence, $m_{qi}$ and $k_{qi}$ cannot be fit to any level curve of these functions, and so a simultaneous big bang is incompatible with both $k_{qi}\to 0$ and $m_{qi}\to 0$ as $r\to\infty$. Since this radial asymptotic behavior of initial value functions comprises models asymptotic to Minkowski in generalized Milne coordinates, self--similar dust solutions and Schwarzschild--Kruskal, models with a simultaneous big bang cannot be asymptotic to any of these spacetimes. This is not surprising since $c\tbb\to -\infty$ holds in all these cases.   
\end{itemize}
\begin{figure}[htbp]
\begin{center}
\includegraphics[width=4in]{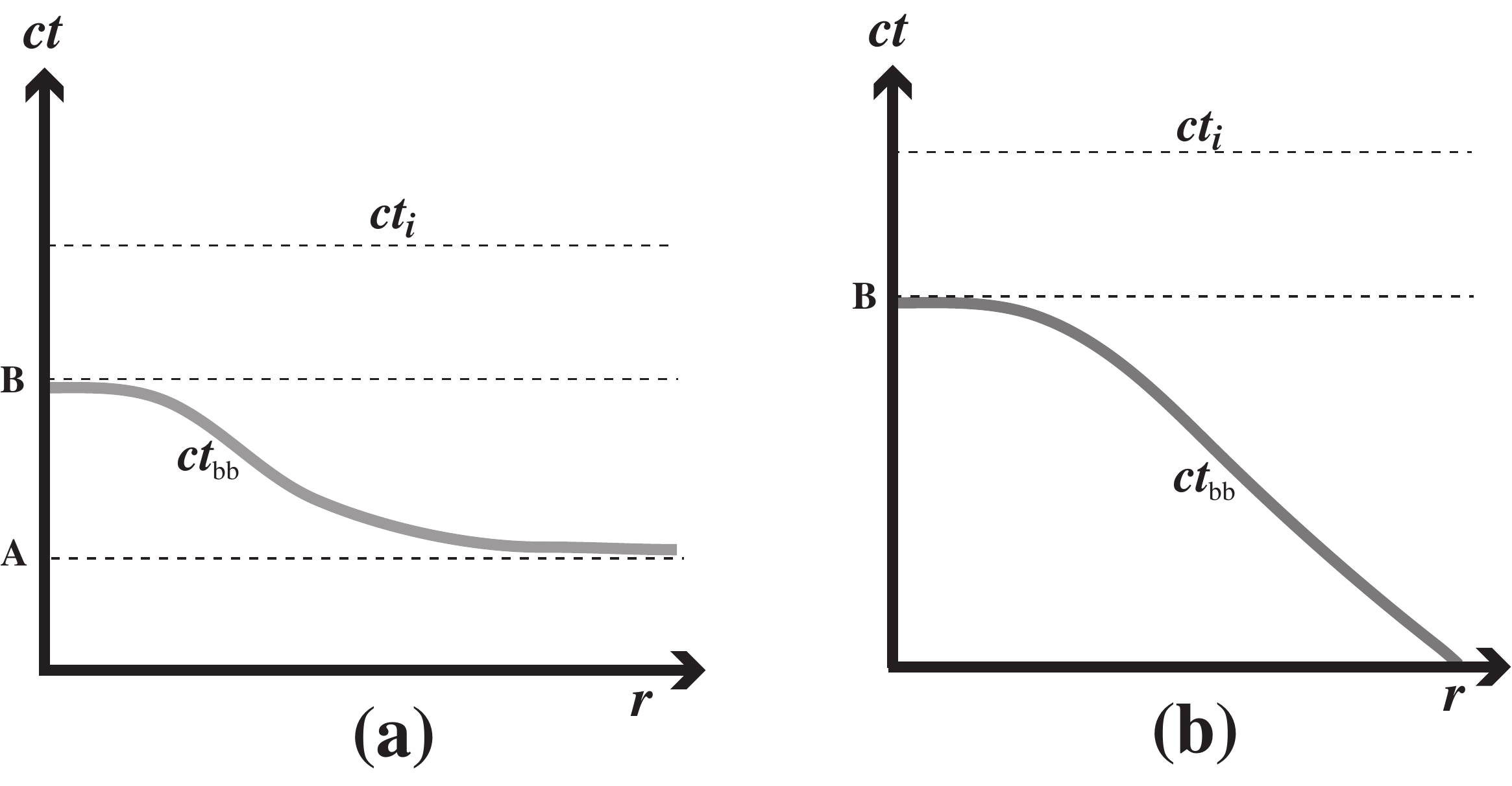}
\caption{{\bf Regularity domain of parabolic and hyperbolic models.} The figure displays the $(ct,r)$ plane with the locus of the initial curvature singularity (big bang) $L=0$, marked by the thick curve $[c\tbb(r),r]$, together with the initial slice $\T[t_i]$ (dotted line). For all models asymptotic to a FLRW or Milne spacetimes (panel (a)), $c\tbb$ tends asymptotically to a constant, while models asymptotic to Minkowski in generalized Milne coordinates, self--similar solutions or Schwarzschild--Kruskal, we have $c\tbb\to-\infty$ in this limit. Notice that space slices $\T[t]$ are fully regular only for $ct>B$.}
\label{fig1}
\end{center}
\end{figure}
\begin{figure}[htbp]
\begin{center}
\includegraphics[width=4in]{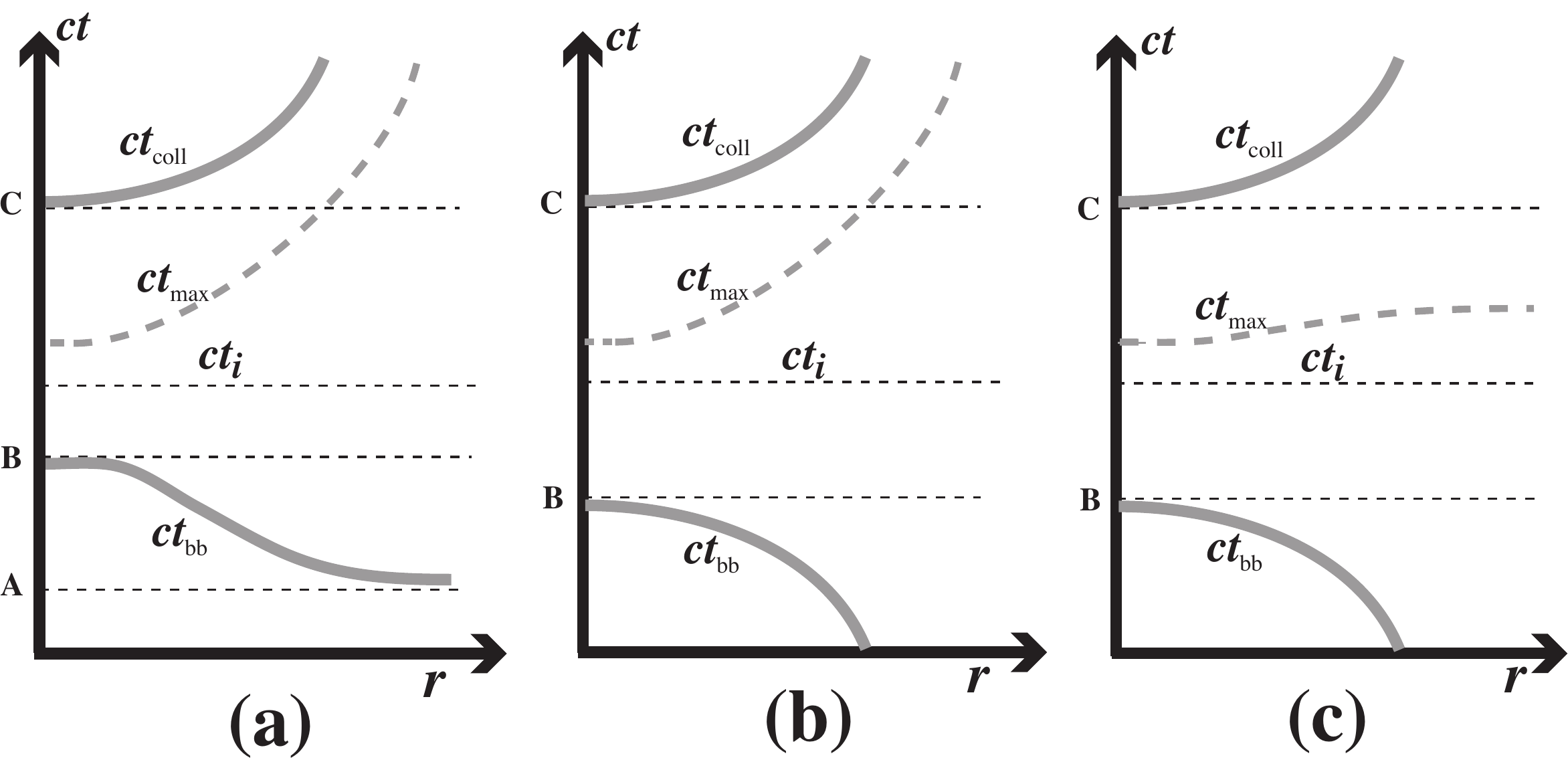}
\caption{{\bf Regularity domain of open elliptic models.} The figure displays the $(ct,r)$ plane with the locus of the initial and collapse curvature singularities corresponding to $L=0$, marked by the thick curves $[c\tbb(r),r]$ and $[c\tcoll(r),r]$, together with the initial slice $\T[t_i]$ (dotted line) and the maximal expansion time ($\HH_q=0$) marked by the thick dotted curve $[c\tmax(r),r]$. Panel (a) corresponds to models asymptotic to the spatially flat FLRW cosmology, so that $c\tbb$ tends asymptotically to a constant (while it tends to $-\infty$ in all other cases). Panel (c) corresponds to the case $k_{qi}/m_{qi}\to 2$ examined in section 11.5, for which $c\tmax$ tends asymptotically to a constant. Panel (b) corresponds to all other cases. For the collapse time we have $c\tcoll\to\infty$ in all cases. Notice that in cases (a) and (b) the full asymptotic radial range is contained in the expanding phase ($ct<c\tmax$). Also, the space slices $\T[t]$ are fully regular only for $B<ct<C$.}
\label{fig2}
\end{center}
\end{figure}
\begin{figure}[htbp]
\begin{center}
\includegraphics[width=3in]{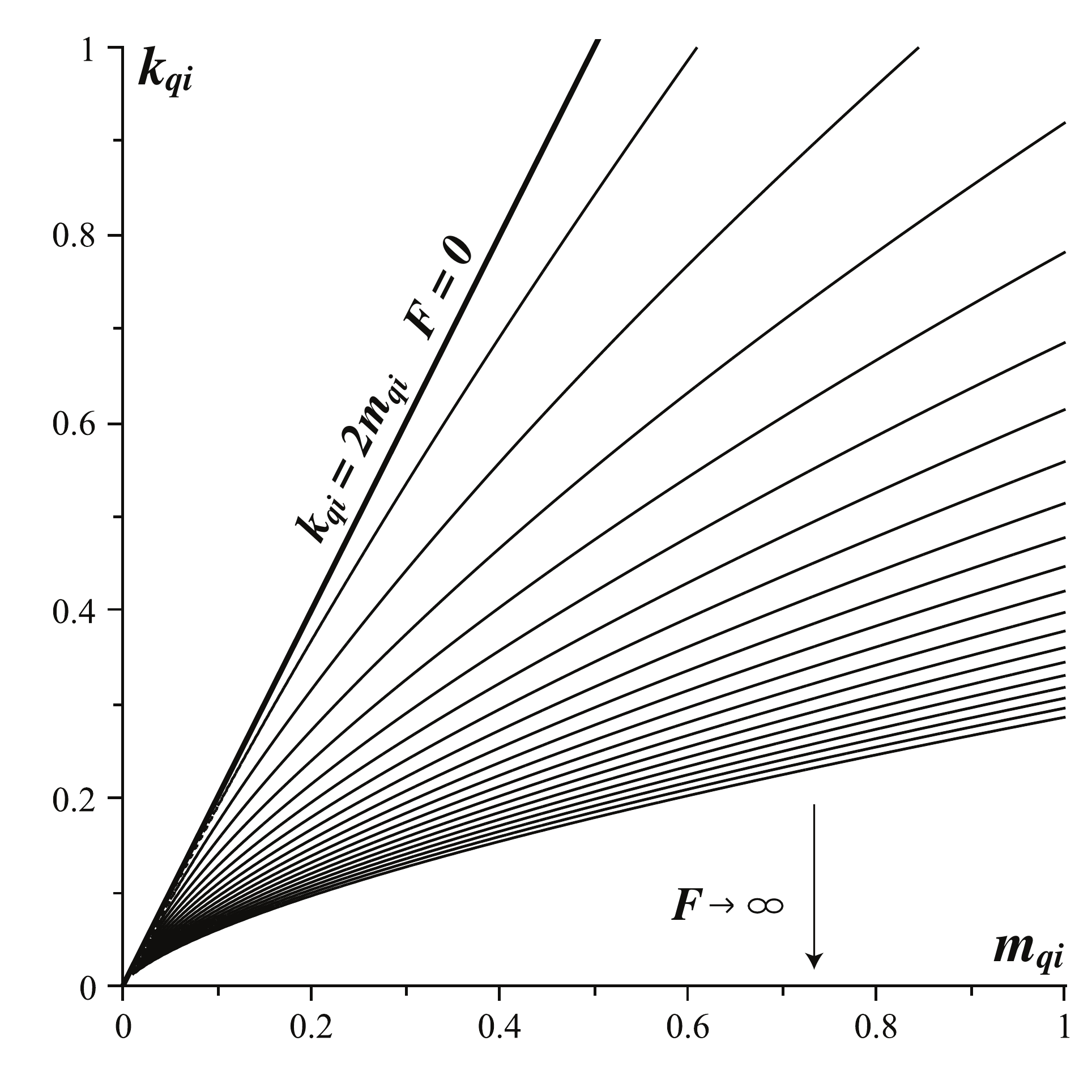}
\caption{{\bf Level curves of $F$ in (\ref{tmce}).} The figure displays the level curves in the $(k_{qi},m_{qi})$ plane. We have $F=0$ in the line $k_{qi} =2m_{qi}$, while $F\to\infty$ as $k_{qi}\to 0$ with $m_{qi}> 0$, while all curves converge at $(k_{qi},m_{qi})\to (0,0)$. As a consequence, only the asymptotic convergence $k_{qi}\to 0$ and $m_{qi}\to 0$ is possible, and so elliptic models with a simultaneous maximal expansion cannot be asymptotic to a FLRW cosmology.}
\label{fig3}
\end{center}
\end{figure}

\subsection{Simultaneous maximal expansion.}  

Let $\tmax=t_{(m)}$ denote the constant time value associated with $\HH_q=0$, then (\ref{tmc}) implies that $m_{qi}$ and $k_{qi}$ are necessarily linked by the constraint
\begin{equation}c(t_{(m)}-t_i) = F(k_{qi},m_{qi})\equiv \frac{m_{qi}\,[\pi-Z_e(k_{qi}/m_{qi})]}{k_{qi}^{3/2}},\label{tmce}\end{equation}
where $Z_e$ follows from (\ref{ellZ1a}) and we have expressed $x_i,\,y_i$ in terms of $m_{qi}$ and $k_{qi}$ by means of (\ref{phie}). This constraint is very similar to (\ref{tbbce}), hence the compatibility with assumptions on radial convergence of $m_{qi}$ and $k_{qi}$ follows from a qualitative study of the level curves of the function $F$ in (\ref{tmce}). 
As shown in figure 3, the level curves of this function do not allow for $k_{qi}\to 0$ to occur with $m_{qi}\to m_0>0$, since $k_{qi}\to 0$ necessarily implies $m_{qi}\to 0$. Therefore, models with in which $c\tmax$ is constant must exhibit the same asymptotic states as the models examined in section 11.5 ($x_i=k_{qi}/m_{qi}\to 2$), in which $c\tmax$ tends to a constant as $r\to \infty$. In fact, the models of section 11.5 asymptotically converge to models with a simultaneous $\tmax$. Depending on the power law decay of $m_{qi}$ and $k_{qi}$, these  models are all asymptotic to Minkowski in generalized Milne coordinates, or to self--similar or Schwarzschild--Kruskal spacetimes.       

\section{Summary, conclusion and final discussion.}    

We have examined the asymptotic regime along the radial direction for regular parabolic, hyperbolic and open LTB models admitting a symmetry center and a covariant time slicing that defines a one--parameter family of 3--dimensional hypersurfaces, $\T[t]$, orthogonal to the 4--velocity and marked by constant $t$. With the help of suitably defined covariant quasi--local scalars, $m_q,\,k_q,\,\HH_q$ and their relative fluctuations, this time slicing allows us to study LTB models under a well posed initial value framework in terms of a fiducial initial hypersurface $\T[t_i]$ marked by $t=t_i$. Since radial rays in the $\T[t]$ are spacelike geodesics of the LTB metric whose affine parameter is proper length, $\ell$, asymptotic conditions naturally correspond to the limit $\ell\to\infty$, which is related to the limit $R\to\infty$ along the radial rays (see section 5). However, $\ell=\ell(t,r)$ and $R=R(t,r)$, and thus it is very difficult to use these invariant quantities to probe the behavior of covariant scalars in this asymptotic limit. Hence, we have provided in section 4 the appropriate conditions in which $\ell\to\infty$ corresponds to the limit $r\to\infty$, so that a regular asymptotic regime follows by demanding that covariant scalars (local and quasi--local) remain bounded in this limit. The asymptotic limit along radial rays then follows from looking at the uniform convergence of initial value functions $m_{qi}(r),\,k_{qi}(r)$ to analytic trial functions $\tilde m_{qi}(r),\,\tilde k_{qi}(r)$ as $r\to\infty$, under the assumption of the radial coordinate gauge (\ref{Rgauge}), which fixes the radial coordinate as proportional to $R_i=R(t_i,r)$ (sections 6 and 7).

\subsection{Classification in terms of their asymptotic limits and states.}  

By assuming asymptotic convergence forms $\tilde m_{qi},\,\tilde k_{qi}$ for $m_{qi},\,k_{qi}$, we examined separately in sections 9--12 the radial asymptotic behavior of the metric functions and the covariant scalars, for parabolic, hyperbolic, open elliptic models and models with a simultaneous big bang or maximal expansion, making the distinction between the ``asymptotic limit'' as the strict limit as $r\to\infty$, and the ``asymptotic state'' as the set of asymptotic expansions of all scalars around the asymptotic limit up to leading terms on $\tilde m_{qi}(r),\,\tilde k_{qi}(r)$ (see section 8). The summary of the results is as follows:\\

\noindent
{\bf{Asymptotic Limit.}} All open LTB models can be classified in terms of two broad classes: asymptotic limit to FLRW or Minkowski. Considering $A=m,\,\HH,\,k$ and $A_q=m_q,\,\HH_q,\,k_q$, the ``conventional'' variables $M,\,E$ and $c\tbb$ (bang time), as well as $c\tcoll,\,c\tmax$ (collapse and maximal expansion times, see (\ref{tmc})), together with the metric functions $L$ and $\Gamma$ in (\ref{LTB2}), and using the radial coordinate gauge (\ref{Rgauge}), we have:

\begin{itemize}

\item Models whose asymptotic limit is a FLRW cosmology, including Milne Universe as a particular vacuum case (see figures 1a and 2a):

\bse\ba \fl L\to \tilde L(t),\qquad R\to R_0\,r\,\tilde L(t)\to\infty,\qquad \Gamma\to 1,\\
\fl A_q \to \tilde A_q(t),\qquad A\to \tilde A_q(t),\qquad\qquad \Da\to 0,\\
\fl M\sim m_0 R_0^3 r^3,\qquad E\sim \pm k_0 R_0^2 r^2, \qquad c\tbb\to \hbox{const.},\\
\fl c\tcoll\to\infty,\qquad c\tmax\to\infty\qquad (\hbox{only elliptic models}).\ea\ese

\item Models whose asymptotic limit is Minkowski, containing self--similar, Schwarzschild--Kruskal and generalized Milne asymptotic states (see figures 1b and 2b--2c):

\bse\ba \fl L\to 1,\qquad R\to R_0\,r\to\infty,\qquad \Gamma\to 1,\\
\fl A_q \to 0,\qquad A\sim A_q(1+\Da)\to 0,\qquad\Da\to \tilde\Da\ne 0\;\;(\hbox{in general}),\\
\fl M\sim \tilde m_{qi}r^3R_0^3,\qquad E\sim \tilde k_{qi} r^2 R_0^2, \qquad c\tbb\to -\infty,\\
\fl c\tcoll\to\infty,\qquad c\tmax\to\infty\quad (\hbox{or}\;\to \hbox{const})\qquad (\hbox{only elliptic models}).\label{ctmaxcons}\ea\ese

\end{itemize}
where the special case $c\tmax \to$ const. in (\ref{ctmaxcons}) has been discussed in sections 11.5 and 12.2.    

\bigskip

\noindent {\bf{Asymptotic States.}} It is useful to list these states for each kinematic class: \\

\noindent {\bf{Parabolic models}} (section 9) are radially asymptotic to

\begin{itemize}

\item FLRW dust with zero spatial curvature (section 9.1)
\item Minkowski in generalized Milne coordinates (section 9.2)
\item Self--similar dust solution with zero spatial curvature (section 9.2)
\item Schwarzschild--Kuskal in Lema\^\i tre coordinates, built by radial geodesics with zero binding energy (section 9.2)

\end{itemize}

\noindent {\bf{Hyperbolic models}} (section 10) are radially asymptotic to

\begin{itemize}

\item A subclass of parabolic models (section 10.3), which are radially asymptotic to

  \begin{itemize}

   \item FLRW dust with zero spatial curvature
   \item Minkowski in generalized Milne coordinates

  \end{itemize}

\item Milne Universe (section 10.4)
\item FLRW dust with negative spatial curvature (section 10.5)
\item Minkowski in generalized Milne coordinates (sections 10.4 and 10.5)
\item Self--similar dust solution with negative spatial curvature (section 10.5)
\item Schwarzschild--Kuskal in coordinates given by radial geodesics with positive binding energy (section 10.4)

\end{itemize}

\noindent {\bf{Open elliptic models}} (section 11) are radially asymptotic to

\begin{itemize}

\item A subclass of parabolic models (section 11.3), which are radially asymptotic to

    \begin{itemize}

    \item FLRW dust with zero spatial curvature
    \item Minkowski in generalized Milne coordinates

     \end{itemize}

\item Minkowski in generalized Milne coordinates (sections 11.4 and 11.5)
\item Self--similar dust solution with positive spatial curvature (sections 11.4 and 11.5)
\item Schwarzschild--Kuskal in Novikov coordinates, built by radial geodesics with negative binding energy (sections 11.4 and 11.5)

\end{itemize}

\noindent {\bf{Models with a simultaneous big bang}} (hyperbolic and open elliptic)  (section 12.1) are radially asymptotic to

\begin{itemize}

\item FLRW dust with zero spatial curvature (as in sections 10.3, 10.5 and 11.3)
\item FLRW dust with negative spatial curvature (as in section 10.5)
\item Milne Universe (as in section 10.4) 

\end{itemize}

\noindent {\bf {Models with a simultaneous maximal expansion}} (elliptic)  (section 12.2) are radially asymptotic to the same spacetimes as those of section 11.5.

\subsection{Cosmological considerations vs radial asymptotics.}

It is usual to consider any given LTB configuration as a model of a spherical inhomogeneity somehow immersed in a cosmic ``background''. In order to discuss this issue in terms of the results obtained in this article, it is illustrative to relate the asymptotic forms of LTB scalars with observational parameters of a FLRW cosmology. 

If we denote by $\Omega(t)$ the FLRW Omega parameter, a possible generalization for LTB models can be given in terms of quasi--local scalars:
\begin{equation} \hOm \equiv \frac{\kappa\,\rho_q}{3\HH_q^2}=\frac{2m_q}{\HH_q^2}=\frac{2m_q}{2m_q-k_q}=\frac{2m_{qi}}{2m_{qi}-Lk_{qi}},\end{equation}
so that each kinematic class (parabolic, hyperbolic or elliptic) follows from the sign of $\hOm-1$
\begin{equation} \hOm-1=\frac{k_q}{\HH_q^2}=\frac{k_{qi} L}{2m_{qi}-Lk_{qi}}. \end{equation}
While $\hOm$ is a covariant quantity (because $m_q$ and $\HH_q$ are covariant), it is not a quasi--local scalar: notice that we do not use the symbol $\Omega_q$ because $\hOm$ cannot be obtained from applying (\ref{aveq_def}) to the ratio $2m/\HH^2$. In fact, other expressions for the Omega and Hubble factors for LTB models have been suggested in the literature \cite{moftar,HMM} (see \cite{suss10} for further discussion on this issue). However, it is evident that for all LTB models whose radial asymptotic limits and states are FLRW cosmologies we have as $r\to\infty$
\begin{equation} \hOm \sim \Omega(t), \end{equation}
which clearly suggests that LTB models with this radial asymptotic behavior are suitable to model cosmological inhomogeneities asymptotically converging to a FLRW ``background''. This means that, irrespective of how we choose to generalize the observational parameters, they will tend asymptotically in the radial direction to the FLRW parameters. Notice that we have reached this conclusion by looking at the asymptotic states of LTB models defined all along the radial rays, without considering the rather artificial situation in which a FLRW  radial asymptotic state of an LTB model is forced by matching to it to a FLRW cosmology at a finite fixed comoving boundary $r=r_b$. Though such configurations can also be constructed with LTB models, leading to smooth and fully relativistic generalizations of the Newtonian ``spherical collapse'' or ``top hat'' models used in qualitative studies of structure formation (see \cite{suss09} for discussion and references on this type of models and their relativistic generalizations). 

The cosmological interpretation of the radial asymptotics of LTB models with asymptotic limit to Minkowski is more complicated. To discuss these cases it is convenient to introduce a ``local' equivalent of $\hOm$ defined with the local scalars $m=\kappa\rho/3$ and $\HH=\Theta/3$ as
\begin{equation}\fl\hOm_{\rm{loc}}\equiv \frac{\kappa\rho}{3\HH^2}=\frac{2m}{2m-k+(\HH_q\Dh)^2}=1+\frac{(\hOm-1)\,(1+\Dk)-(\Dh)^2}{(1+\Dh)^2},\end{equation}
where we eliminated $\HH$ and $k$ from (\ref{qltransf}) and used the constraint
\begin{equation}  \HH^2 = 2m-k+\Sigma^2=2m-k+(\HH_q\Dh)^2, \end{equation}
which follows from the definitions of $\Theta$ and $\Sigma$ in (\ref{ThetaRR}) and (\ref{SigEE1}). For models with an asymptotic FLRW limit, $\Dh\to 0$ and $\Dk\to 0$ hold, so $\hOm_{\rm{loc}}\to \hOm$ and the FLRW $\Omega(t)$ results. However, in models with a Minkowski limit $\Dh,\,\Dk$ do not tend to zero, resulting in a different form for $\hOm_{\rm{loc}}$.  

In parabolic models with asymptotic limit to Minkwoski we have $\hOm =1$ everywhere, while $\Dh\to\Dih=\Dim/2$ (from (\ref{asPar3d})), hence
\begin{equation} \hOm_{\rm{loc}}\approx 1-\left(\frac{\Dih}{1-\Dih}\right)^2 \leq 1,\label{Omlocp}\end{equation}
Since we have $\hOm\to 1$ for hyperbolic with a matter dominated Minkowski limit (section 10.4) and for all open elliptic models with a Minkowski limit (which is also matter dominated), we obtain for these cases the same limit as in (\ref{Omlocp}). For all hyperbolic models with a vacuum dominated Minkowski limit we have $\hOm\to 0$, hence
\begin{equation} \hOm_{\rm{loc}}\approx \frac{2\Dh-\Dk}{(1+\Dh)^2}.\end{equation}
Therefore, considering (\ref{asHyp23d})--(\ref{asHyp23e}), (\ref{asHyp33c})--(\ref{asHyp33d}) and (\ref{asHyp54c})--(\ref{asHyp54d}), we have $\Dk\to 2\Dh$, and so $\hOm_{\rm{loc}}\to 0$ holds for models asymptotic to Milne, generalized Milne and self--similar solutions, but in models asymptotic to Schwarzschild--Kruskal we have $\Dk\to \Dh$ (see  (\ref{asHyp43d})--(\ref{asHyp43e})), so for these models $\hOm_{\rm{loc}}$ has an asymptotic form as in (\ref{Omlocp}).   

Since the asymptotic forms for either $\hOm$ or $\hOm_{\rm{loc}}$ are independent of $t$ (being either zero or a constant $\leq 1$), a cosmological interpretation for the radial asymptotics of models with a Minkowski limit cannot be given in terms of a smooth convergence  or transition to a FLRW model in the radial direction. Instead of a cosmic FLRW background, the external realm of these LTB models could be a large void region. A similar and appealing interpretation of radial asymptotics to Minkowski follows naturally from the notion of ``finite infinite'' (fi), introduced by Ellis \cite{fiEllis} (see also comments in \cite{fiWilt}) to represent a near Minkowskian timelike envelope around bound structures, which defines a scale in which these structures can be studied as almost asymptotically flat systems without considering the effect of cosmic expansion. Such a scale would be intermediate between characteristic lengths of bound structures and a cosmic scale where expansion cannot be ignored. Evidently, LTB models with asymptotic limit to Minkowski approximate spherically symmetric realizations of this cosmic structure in such a near Minkowski envelope.          

The results of this article are useful for probing supernovae and CMB observations by means of LTB models. They are essential for the study of radial profiles of covariant scalars, namely: the ``clump'' or ``void'' profiles and the possibility of profile inversions as the models evolve in time. This study has been undertaken in a separate article \cite{rprofs}. The radial asymptotic properties of the models have also theoretical and practical consequences in the application of Buchert's scalar averaging formalism \cite{ave_review} to LTB models, as vacuum dominated LTB  models are the LTB configurations most likely to yield an ``effective'' acceleration that mimics dark energy emerging from back--reaction terms in the context this formalism. These connections have been already remarked \cite{LTBave2,LTBave3} and are currently under further investigation.  

\begin{appendix}

\section{Analytic solutions in the conventional variables.}

The solutions of the Friedman--like equation (\ref{fieldeq1}) in terms of the conventinal free functions $M,\,E,\,c\tbb$ are:

\begin{itemize}
\item{Parabolic models or regions: \,\, $E=0$.}
\begin{equation}c(t-\tbb) = \frac{2}{3}\,\eta^3,\qquad R=(2M)^{1/3}\,\eta^2,\label{par1}\end{equation}
\item{Hyperbolic models or regions: \,\, $E\geq 0$.}
\ba
\fl R =\frac{M}{E}\,\left(\cosh\,\eta-1\right),\qquad c(t-\tbb)=\frac{M}{E^{3/2}}\,\left(\sinh\,\eta-\eta\right)\label{hypRt}\\
\fl \frac{E^{3/2}}{M}\,c(t_i-\tbb)=Z_h(U),\qquad U\equiv \frac{ER}{M},\label{hyptR}
\ea
\item{Elliptic models or regions: \,\, $E\leq 0$.}
\ba
\fl R =\frac{M}{|E|}\,\left(1-\cos\,\eta\right),\qquad c(t-\tbb)=\frac{M}{|E|^{3/2}}\,\left(\eta-\sin\,\eta\right),\label{ellRt}\\
\fl \frac{|E|^{3/2}}{M}\,c(t_i-\tbb)=\left\{ \begin{array}{l}
  Z_e(U)\quad \hbox{expanding phase}\quad \dot R> 0 \\ 
  2\pi-Z_e(U) \quad \hbox{collapsing phase}\quad \dot R< 0 \\ 
 \end{array} \right.,\quad  U\equiv \frac{|E|R}{M},\nonumber\\\label{elltR}           
\ea
\end{itemize}
where $Z_h$ and $Z_e$ are respectively given by (\ref{hypZ1a}) and (\ref{ellZ1a}). The solutions presented above are given in terms of the new variables by (\ref{par2}), (\ref{phi_L_h}) and (\ref{ellZ2}).

Another particular solution of (\ref{fieldeq1}) follows by assuming $M=0$ and $E>0$ (but otherwise arbitrary), leading to the case ``[s2]'' in \cite{ltbstuff}
\begin{equation} R = E\,c(t-\tbb),\label{s2case}\end{equation}
which are locally Minkowski solutions (sections of Minkowski spacetime in coordinates that generalize Milne's Universe).  

\section{Prescribing local initial value functions.}

In section 7 we assumed an asymptotic convergence for for $A_{qi}$ and obtained the forms for $A_i$ and $\Da_i$. We examine here the opposite situation.  If we assume that $A_i\sim \tilde A_i$, then for $r>y$ and with the help from (\ref{aveq_def}) and (\ref{Rgauge}) we have
\begin{equation} A_{qi}=\frac{3}{r^3}\int_0^r{A_i x^2 \dd x}\sim \frac{3}{r^3}\left[\int_0^y{A_i x^2 \dd x}+\int_y^r{\tilde A_i x^2 \dd x}\right],\label{Aqsim}\end{equation}
while $\Da_i$ follows directly from (\ref{Dadef}):
\begin{equation} \Da_i \sim \frac{\tilde A_i(r) r^3-3\int_0^y{A_i x^2 \dd x}-3\int_y^r{\tilde A_i x^2  \dd x}}{3\int_0^y{A_i x^2 \dd x}+3\int_y^r{\tilde A_i x^2 \dd x}}.\label{Daas}\end{equation}
A comparison of (\ref{Asim})--(\ref{Daasq}) and (\ref{Aqsim})--(\ref{Daas})shows that it is much easier to obtain the asymptotic limit of $\Da_i$ from $A_{qi}$ than from $A_i$. Prescribing $A_{qi}$ is also more useful in practice, since the initial value functions in the analytic solutions given in sections 9--11 are $m_{qi},\,k_{qi}$.

We now examine the asymptotic forms for $m_{qi}$ and $\Dim$ that follow given a prescribed asymptotic form for $m_i$. Consider a power law decay $m_i\sim m_0 r^{-\alpha}$, then (\ref{Aqsim}) and (\ref{Daas}) yield
\ba\fl   m_{qi}\sim m_0\left[\frac{I(y)}{r^3}+\frac{3}{(3-\alpha)\,r^\alpha}\right],\qquad \Dim\sim \frac{-\alpha\,r^{-\alpha}-(3-\alpha)I(y)r^{-3}}{(3-\alpha)I(y)r^{-3}+3\,r^{-\alpha}},\label{AqDa}\\
\hbox{where:}\qquad I(y)=\frac{3}{m_0}\int_0^y{m_i x^2 \dd x}-\frac{3}{3-\alpha}y^{3-\alpha},\label{Idef}\nonumber\ea
Notice that if $\alpha<3$ then $r^{-\alpha}$ is the dominant term, whereas if $\alpha>3$ then $r^{-3}$ is dominant. Hence, we obtain $m_{qi}\propto r^{-\alpha}$ for $\alpha<3$ and for a logarithmic decay, while $m_{qi}\propto r^{-3}$ results if we assume that  $\alpha\geq 3$ or an exponential decay. However, from Lemma 3, both $m_i$ and $m_{qi}$ tend to zero. By assuming $r\gg y$ we obtain the following asymptotic limit for $\Dim$
\begin{equation} \Dim \sim \tilde\Dim =\left\{ \begin{array}{l}
  - \alpha /3,\quad \hbox{if}\quad \alpha  < 3 \\ 
  - 1\quad \quad \;\;\hbox{if}\quad \alpha  \ge 3 \\ 
 \end{array} \right.,\label{Da00}\end{equation}
It is straightforward to show that if $m_i$ decays to zero exponentially, then $\tilde\Dim\to-1$, while a slow logarithmic decay yields $\tilde\Dim\to 0$. As opposed to prescribing an asymptotic decay for $m_{qi}\to 0$ as in (\ref{decmq2})--(\ref{decmq3}), all assumptions on the decay of $m_i\to 0$ necessarily yield $m_i$ and $m_{qi}$ positive in the full range $r>y$. This is so because $m_{qi}\geq m_i\geq 0$ holds in this range (as $m'_i$ and $m'_{qi}$ are negative, see (\ref{propq2})).

\section{Particular cases of open LTB models.}

Given the radial coordinate gauge (\ref{Rgauge}), all open LTB models can be univocally characterized by the initial value functions $m_{qi},\,k_{qi}$. Hence, spacetimes that emerge as particular cases of these models emerge by specific specialization of these functions. The line elements of these spacetimes follow directly from (\ref{LTB2}) by recalling the particular cases of these functions accordingly, and define particular case spacetimes which have served as references to the possible asymptotic states of all open LTB models in the limit $r\to\infty$.     

\subsection{FLRW dust spacetimes} 

Homogeneous and isotropic dust FLRW cosmologies follow as particular cases of LTB models by setting 
\begin{equation}  m_{qi} = m_0>0\quad\hbox{and}
\quad k_{qi}=k_0,\label{FLRW}\end{equation}
where $m_0,\,k_0$ are constants. From (\ref{ME}), (\ref{tbbpar}), (\ref{tbbh}) and (\ref{tmc}), we have
\begin{equation} \fl M=m_0\,r^3,\quad E=-k_0\,r^2,\quad \tbb,\,\tmax,\,\tcoll\quad\hbox{constants}.\label{MEtbb1}\end{equation}
Evidently, the spatially flat case corresponds to $k_0=0$, while positive/negative $k_0$ are FLRW models with positive/negative spatial curvature. It is evident from (\ref{par2}) that (\ref{FLRW}) yields for $k_0=0$
\begin{equation} L=L(t) = \left[1+\frac{3}{2}\sqrt{2m_0}\,c(t-t_i)\right]^{2/3},\qquad k_0=0,\label{FLRW_L0}\end{equation}
while for $k_0\ne 0$ the function $L=L(t)$ follows from the implicit solutions (\ref{phi_L_h}) and (\ref{ellZ2}) under the specialization (\ref{FLRW}):
\begin{equation} \fl \phi(t)=y_0\,c(t-t_i)+Z(x_0) = Z(x_0\,L), \qquad x_0=\frac{|k_0|}{m_0},\quad y_0=\frac{|k_0|^{3/2}}{m_0},\label{FLRW_L}\end{equation}
with $Z=Z_h$ or $Z=Z_e$ given by (\ref{hypZ1a}) and (\ref{ellZ1a}) for hyperbolic (negative curvature) and elliptic (positive curvature) FLRW models. For all particular cases (\ref{FLRW}), equations (\ref{Dadef}) and (\ref{qltransf}) imply $\Dim=\Dik=\Dih=0$, so that $m_i=m_{qi},\,k_i=k_{qi}$ and $\HH_i=\HH_{qi}$ hold. Also, (\ref{Hq}), (\ref{Gp}) and (\ref{Ghe}) imply $\Gamma=1$, and thus (\ref{slawDm}) and (\ref{slawDk}) yield $\Dm=\Dk=0$, and so, $\Dh=0,\,m=m_q,\,k=k_q$ and $\HH=\HH_q$ follow from (\ref{slaw1}), (\ref{slaw2}) and (\ref{slawDh}). As a consequence, (\ref{LTB2}) becomes a FLRW line element and all scalars are either constants or functions of $t$ only.  

\subsection{Minkowskian particular cases.}

Considering the Riemann tensor (see Appendix A2 of \cite{suss10}), any particular LTB model in which $m_q=m=0$ is locally equivalent to Minkowski spacetime given in curvilinear coordinates that generalize the well known Milne Universe. These are the locally Minkowskian solutions (\ref{s2case}), corresponding to the case ``[s2]'' in the classification of \cite{ltbstuff}, and are given by the specialization:
\begin{equation} m_{qi}=0 \quad\hbox{and}\quad  k_{qi}<0\quad\hbox{but otherwise arbitrary.}\label{Mink}\end{equation}
so that $m_q=m=\Dm=0$, but $\Dk\ne 0,\,\Dh\ne 0$. The metric functions $L$ and $\Gamma$ follow from substituting (\ref{Mink}) into the equation for $\dot L$ in (\ref{Hq}) and substituting in the definition of $\Gamma$ in (\ref{LGdef}):
\begin{equation}\fl  L = 1+|k_{qi}|^{1/2}\,c(t-t_i),\qquad \Gamma= \frac{1+|k_{qi}|^{1/2}\left(1+\frac{3}{2}\Dik\right)\,c(t-t_i)}{1+|k_{qi}|^{1/2}\,c(t-t_i)}\label{Mink_L}\end{equation}
where we used (\ref{Dadef}) and the coordinate gauge (\ref{Rgauge}). Evidently, we have $M=0$, while $E$ is arbitrary as in (\ref{s2case}). The scalars associated with spatial curvature and expansion correspond to hypersurfaces $\T$ and a 4--velocity associated with test observers. In general, these coordinates do not cover the full Minkowski spacetime, and thus $L=0$, marked by $ct=c\tbb(r)=ct_i-1/|k_{qi}|^{1/2}$, corresponds to a caustic surface for this congruence, not to a curvature singularity. Milne's Universe, which can also be considered as a vacuum FLRW cosmology, is the particular case of (\ref{Mink}) and (\ref{Mink_L}) given by 
\ba  m_{qi}=0 \quad\hbox{and}\quad  |k_{qi}|=k_0=\hbox{constant},\label{Milne}\\
 L = L(t)= 1+\sqrt{k_0}\,c(t-t_i),\label{Milne_L}\\
 M=0,\quad E=\FF^2-1=|k_0|\,r^2,\quad \tbb=\hbox{constant}.\label{MEtbb2}\ea
Since $\Dik=0$ in this case vanish, then $\Gamma=1$ and all the $\Da$ also vanish. Also, we have $m_q=m=0$ for all $t$, though $k_q=k<0$ and $\HH_q=\HH$ are functions of time. In the general case (\ref{Mink}) $k_q=k<0$ and $\HH_q=\HH$ are also functions of $r$.   

\subsection{Schwarzschild--Kruskal spacetime.} 

The Schwarzschild--Kruskal spacetime in comoving coordinates constructed by radial geodesics (Lema\^\i tre and Novikov coordinates, see page 332 of \cite{kras2}) follows by setting
\begin{equation}  m_{qi}=m_0\,r^{-3}\quad\hbox{and}\quad k_{qi}\:\:\hbox{arbitrary},\label{Schw}\end{equation}
Hence, (\ref{ME}) and (\ref{Rgauge}) imply
\begin{equation} M=M_0=m_0\,R_0^3,\qquad E=\FF^2-1\:\:\hbox{arbitrary},\label{MEtbb3}\end{equation}
while (\ref{Dadef}) and (\ref{slawDm}) yield $\Dim=\Dm=-1$, so that $2m_q=2M_0/R^3$ and $m_i=m=0$ follows from (\ref{qltransf}) and (\ref{slaw1}). It is straightforward to prove that the Riemann tensor (see Appendix A2 of \cite{suss10}) with $m=0$ and $m_q\ne 0$ leads to a vanishing Ricci tensor, and thus we can identify this case as a vacuum Schwarzschild--Kruskal solution where $2M_0$ is the Schwarzschild radius. The variables associated with spatial curvature ($k,\,k_q$ and $\Dk$) and the expansion ($\HH,\,\HH_q$ and $\Dh$) remain arbitrary because a spatial curvature and an expansion can always be associated to a congruence of radial geodesic test observers with arbitrary binding energy. In fact, by setting $M=M_0$ in (\ref{fieldeq1}) we obtain the equation for radial timelike geodesics in a Schwarzschild spacetime if we identify $\FF^2-1$ with the binding energy of the comoving test observers.

The form of $L$ depends on the choice of $k_{qi}$. If we choose geodesic observers with zero binding energy $k_{qi}=0$, then $L$ and $\tbb$ follow from (\ref{par2}) and (\ref{tbbpar}) with $m_{qi}$ given by (\ref{Schw}):
\begin{equation}\fl L(t,r)=\left[1+\frac{3}{2}\sqrt{2m_0}\,r^{-3/2}\,c(t-t_i)\right]^{2/3},\qquad c\tbb= ct_i- \frac{2\,r^{3/2}}{3\sqrt{2m_0}}.\label{Schw_L0}\end{equation}
If we choose $k_{qi}\ne 0$, then $L$ follows from the implicit solutions (\ref{phi_L_h}) and (\ref{ellZ2}) under the specialization (\ref{Schw}):
\begin{equation} \fl \phi(t,r)=y_i\,c(t-t_i)+Z(x_i) = Z(x_i\,L), \qquad x_i=\frac{k_{qi}\,r^3}{m_0},\quad y_i=\frac{k_{qi}^{3/2}r^3}{m_0}.\label{Schw_L}\end{equation}
where $Z$ is either one of the functions (\ref{hypZ1a}) or (\ref{ellZ1a}). Notice that $\tbb\to-\infty$ as $r\to\infty$ in (\ref{Schw_L0}), whereas the form of $\tbb$ (and $\tmax,\,\tcoll$ if $k_{qi}> 0$) depend on the choice of $k_{qi}$. As shown in sections 10 and 11, if $k_{qi}^{3/2}r^3\to 0$, then $L$ tends asymptotically to (\ref{Schw_L0}). We also remark that the Schwarzschild--Kruskal spacetimer lacks a regular symmetry center, hence $r=0$ is not a special worldline. Also, the comoving coordinates in the cases $k_{qi}\leq 0$ do not cover the full Schwarzschild-Kruskal manifold.   

\subsection{Self--similar solutions.}

The self--similar LTB solutions have been popular models to study gravitational collapse \cite{sscoll}. They have been classified in \cite{selfsim}. In terms of the parameters used in this article these  solutions follow as the particular case
\begin{equation}  m_{qi}=m_0\,r^{-2}\quad\hbox{and}\quad k_{qi}=k_0\,r^{-2},\label{SS}\end{equation}
where $m_0>0$ and $k_0$ are constants. The form of $L$ follows from (\ref{par2}) and from the implicit solutions (\ref{phi_L_h}) and (\ref{ellZ2}) under the specialization (\ref{SS}):
\ba\fl L=\left[1+\frac{3}{2}\sqrt{2m_0}\,\zeta\right]^{2/3},\qquad k_0=0\label{SS_L0}\\
\fl \phi(t,r)=y_0\,\zeta+Z(x_0) = Z(x_0\,L), \qquad x_0=\frac{k_0}{m_0},\quad y_0=\frac{k_0^{3/2}}{m_0},\quad k_0\ne 0\label{SS_L}\ea
where $Z=Z_h$ or $Z=Z_e$ given by (\ref{hypZ1a}) and (\ref{ellZ1a}) for hyperbolic and elliptic models, and we can identify the self similar variable 
\begin{equation} \zeta \equiv \frac{c(t-t_i)}{r}.\label{SSvar}\end{equation}
The functions $M,\,E$ and $\tbb$ follow from applying (\ref{SS}) to (\ref{ME}), (\ref{tbbpar}) and (\ref{tbbh}) 
\ba \fl M=m_0 R_0^3\,r,\qquad E=\FF^2-1 =-k_0\,R_0^2,\label{MEtbb41}\\
\fl c\tbb=ct_i-\frac{2\,r}{3\,\sqrt{2m_0}},\quad k_0=0,\qquad c\tbb=ct_i-\frac{r}{y_0}\,Z(x_0),\quad k_0\ne 0.\label{MEtbb42}  \ea
As in the Schwarzschild--Kruskal case, we have $\tbb\to-\infty$ as $r\to\infty$ (see section 13). As a consequence of (\ref{SS}) we have (in general) $\Dim=\Dik=-2/3$. The forms for $m_q,\,\HH_q,\,k_q,\,\Gamma,\,\Dm,\,\Dk,\,\Dh$ and the remaining scalars follow by applying (\ref{SS}) to the appropriate equations. It is straightforward to verify that the metric (\ref{LTB2}) takes the appropriate self--similar form, admitting the homothetic vector $z^a=c(t-t_i)\delta^a_0+r\delta^a_r$. Also, all dimensionless variables ({\it{i.e.}} of the form $A/A_i$, as well as $L,\,\Gamma$ and $\HH_q\,c(t-t_i)$) depend only on the self--similar variable.

\end{appendix}

\end{document}